\documentclass[12pt,thmsb,notitlepage]{article}
\usepackage{amsmath}
\usepackage{amssymb}
\usepackage{hanging}

\newtheorem{theorem}{Theorem}[section]

\newtheorem{lemma}[theorem]{Lemma}

\newtheorem{ass}{Assumption}
\usepackage{lscape}
\usepackage{booktabs}
\begin{document}

\title{Limit Theorems for Factor Models\thanks{%
We thank the Editor Peter C.B. Phillips, Co-Editor Guido Kuersteiner, and three anonymous referees for guidance and helpful suggestions that have greatly improved the
paper.\smallskip }}
\author{Stanislav Anatolyev\thanks{%
Address: Stanislav Anatolyev, CERGE-EI, Politickych v\v{e}z\v{n}\r{u} 7,
11121 Prague 1, Czech Republic; e-mail \texttt{%
stanislav.anatolyev@cerge-ei.cz}. Czech Science Foundation support under
grants 17-26535S and 20-28055S is gratefully acknowledged.\smallskip } \\
CERGE-EI and NES \and Anna Mikusheva\thanks{%
Address: Department of Economics, M.I.T., 50 Memorial Drive, Building E52,
Cambridge, MA, 02142, USA; e-mail \texttt{amikushe@mit.edu}. National
Science Foundation support under grant 1757199 is gratefully
acknowledged. } \\
MIT\bigskip }
\date{September 2020 \\
}
\maketitle

\begin{abstract}
\quad \newline
\noindent The paper establishes central limit theorems and proposes how to perform valid inference in factor models. We consider a setting where many counties/regions/assets are observed for many time periods, and when estimation of a global parameter includes aggregation of a cross-section of heterogeneous micro-parameters estimated separately for each entity. The central limit theorem applies for quantities involving both cross-sectional and time series aggregation, as well as for quadratic forms in time-aggregated errors. The paper studies the conditions when one can consistently estimate the asymptotic variance, and proposes a bootstrap scheme for cases when one cannot. A small simulation study illustrates performance of the asymptotic and bootstrap procedures. The results are useful for making inferences in two-step estimation procedures related to factor models, as well as in other related contexts. Our treatment avoids structural modeling of cross-sectional dependence but imposes time-series independence.

\bigskip \bigskip

\bigskip

\bigskip
\bigskip \bigskip
\end{abstract}

\thispagestyle{empty} \setcounter{page}{0} \baselineskip=18.0pt

\section{INTRODUCTION}

Data with an underlying factor structure are increasingly used in empirical
macroeconomics and finance. Often these data consist of time series of
observations for multiple cross-sectional units (assets, portfolios, regions
or industries). Quite a few new estimation strategies have appeared in the
empirical literature that use both cross-sectional and time series variation
in order to estimate global structural parameters. Often the parameter of
interest arises from aggregation or estimation using cross-sectional
variation of individual parameters for each entity. One example of such a
structure is linear factor pricing model in asset pricing (Fama \& MacBeth
1973 and Shanken 1992), where for estimation we usually use time series of
excess returns for a number of portfolios or assets priced by a small number
of risk factors. Each portfolio or stock may have its own (heterogeneous)
exposure to risk, often referred to as betas, which can be estimated
separately from time series observations for each portfolio. The parameter
of interest, a risk premium, is defined as the coefficient of
proportionality in the cross-sectional relation between the average excess
return on a portfolio and its individual beta.

A vast majority of macroeconomic shocks are only weakly identified via
structural VARs that use only time series observations on leading macro
variables. A new approach to the estimation of causal effects of a macro
shock on the economy is to use cross-sectional variation in data on regions,
countries or industries. For example, Serrato \& Wingender (2016) use
cross-sectional variation in federal spending programs due to a Census shock
to identify the causal impact of government spending on the economy.
Cross-sectional variation among counties in government spending and in the
accuracy of census-based estimates of population provides a better justified
treatment effect framework, allows for the estimation of local fiscal
multipliers, and finally gives a better global estimate of the fiscal
multiplier via aggregation of local multipliers. Hagedorn et al. (2015)
estimate the aggregate effect of unemployment-benefit duration
on employment and labor force participation using cross-sectional
differences across US states. Sarto (2018) discusses how heterogeneous
sensitivities of regions to aggregate policy variables, so called
micro-global elasticities, can be used to recover macro elasticities of
interest such as, for example, a fiscal multiplier.

A shared feature of the above-mentioned examples is the use of time-series
observations on multiple entities (stocks, portfolios, counties, states or
industries), while data on those entities are not independent and
identically distributed. Moreover, variables for different entities often
display strong co-movements to the extent that the data have a factor
structure, and estimation of these co-movements is the main goal. Indeed,
the realization of a risk factor in the economy moves returns on all
portfolios simultaneously, while a federal fiscal shock moves spending in
all US counties, though in both cases heterogeneously so. A valid estimation
procedure must explicitly model and account for the data's factor structure
to the extent that the error terms (or residuals) can be considered
idiosyncratic; see Kleibergen \& Zhan (2015) and Anatolyev \& Mikusheva
(2018) for how a factor structure that is unaccounted for can lead to
misleading results. However, idiosyncrasy of the errors usually implies
only that the correlation among errors for different entities is relatively
small and does not introduce first-order bias to the estimation procedure.
Usually, it is not reasonable to assume that errors for different entities
are completely independent; indeed, stocks in the same industry are likely
to co-move even after global-economy risks are removed, while errors for
neighboring counties are more likely to be correlated even after one
accounts for federal shocks. At the same time, we typically want to remain
agnostic about the correlation structure of shocks and avoid their
structural modeling as long as this does not introduce biases.

The second typical feature of the above-mentioned examples is the two-step
nature of the estimation procedure, where in the first step we estimate
entity-specific coefficients (risk exposures/betas, local fiscal
multipliers, micro-global elasticities) by running a time-series regression
separately for each entity. In the second step, we estimate the global
coefficient of interest by either aggregating entity-specific coefficients
(Serrato \& Wingender 2016 and Hagedorn et al. 2015), or by
running an OLS regression on the cross-section of entity-specific
coefficients (Fama \& MacBeth 1973 and Sarto 2018), or by running an IV
regression on the cross-section of entity-specific coefficients (Anatolyev
\& Mikusheva 2018).

The goal of this paper is to establish  central limit theorems (CLTs) and to
provide a tool for establishing asymptotic normality of estimates obtained
in such two-step estimation procedures and for finding ways to do
asymptotically correct inference, while being flexible in modeling the
cross-sectional dependence of errors. The main difficulty here is that even
though the second step cross-sectional regression has nearly uncorrelated
errors (which is usually sufficient to obtain consistency of the two-step
estimator), this condition is usually insufficient for a CLT, which
typically requires that stronger discipline be imposed on the dependence
structure (such as independence, or a martingale difference structure, or
mixing). Our solution to this problem is to restrict the time series
behavior while staying agnostic about the cross-sectional dependence. We
assume time-series independence of idiosyncratic errors, which is consistent
with market efficiency for factor asset pricing models and the
non-predictability of macro shocks in macroeconomic settings. The estimation
noise in a two-stage procedure involves aggregation both over time (from the
first step) and over entities (from the second step). We show that under
certain conditions it is sufficient to have a CLT over just one of these
directions, and we use the time-series direction for that.

When the second step uses an OLS or IV estimator, the CLT must adapt to
averages of \textit{quadratic forms}, as both the second-step-dependent
variable and the second stage regressor/instrument contain first-stage
estimation noise. Our CLT has a linear and a quadratic part. We also note
that a need for a CLT for quadratic forms in factor models sometimes arises
for the first-step estimators (e.g., Pesaran \& Yamagata 2018) or in
higher order asymptotic derivations (e.g., Bai \& Ng 2010).

There is a growing literature that establishes different CLTs while
acknowledging the importance of cross-sectional dependence in the data,
which stems from spatial relations and/or from the presence of common
factors. Kuersteiner \& Prucha (2013) establish a CLT for linear sums in a
panel data context with growing cross-sectional dimension $N$ and fixed
time-series dimension $T$ allowing for cross-sectional dependence, and
Kuersteiner \& Prucha (2020) extend these results to quadratic forms as
well. Both papers impose conditional moment restrictions, which allows the
authors to construct a martingale difference sequence in the cross-sectional
direction. The main conditional moment restrictions imply a correct
specification of an underlying model, which need not be required by our CLT.
However, the mentioned papers allow more flexibility in modeling the time
dependence, and do not require large $T$. Another CLT that requires both
large $N$ and large $T$ is established in Hahn et al. (2020) for linear
terms only.

This paper also contributes to the literature on the CLT for quadratic
forms. Various types of CLTs for quadratic forms have been previously
established and used in the many instrument literature (see for example,
Chao et al. 2012, Hausman et al. 2012, S\o lvsten 2020) and many covariate
literature (see Cattaneo et al. 2018), as well as in the literature on
semi-parametric estimation (Cattaneo et al. 2014a, 2014b). The CLT used in
those papers are established for the cross-sectional dimension only, and
rely heavily on the independence assumption. We adapt the ideas used in Chao
et al. (2012), specifically the approach of de Jong (1987), to accommodate
large cross-sectionally dependent panels; an alternative approach, known as
Stein's method, is used in S\o lvsten (2020).

Our second set of results is related to ways of conducting valid statistical
inference. Under strengthened conditions on the weakness of the
cross-sectional correlation of errors, we show that a conventional variance
estimator is consistent, and so the usual asymptotic inference can be
applied. When such strengthened conditions do not hold, we propose instead a
variant of a wild bootstrap scheme that replicates the original
cross-sectional dependence structure. We also conduct a small simulation
experiment that provides evidence on the approximation quality of our CLT
and on the empirical size and power of wild bootstrap in a moderately sized
panel.

The paper proceeds as follows. Section \ref{section-notation} explains
problems with establishing asymptotic Gaussianity for two-step and other
estimators and test statistics, and shows how discipline in the time series
direction can help. Section \ref{section-CLT} introduces assumptions on
idiosyncratic errors, states central limit theorems for two cases, and
discusses the relevance of those cases to empirical practice. Section \ref%
{section-variance} discusses estimation of asymptotic variances for
asymptotic inference and alternative inference tools based on the
bootstrap. Section \ref{section-simulations} presents a small simulation
experiment that reveals properties of asymptotic and proposed bootstrap
inference tools. Section 6 concludes. All proofs appear in the Appendix.

\section{GOALS\ AND\ EXAMPLES}

\label{section-notation}Let the data contain observations on many units
indexed by $i=1,...,N,$ and observed for multiple time periods $t=1,...,T.$
We assume that both $N$ and $T$ increase to infinity without restrictions on
their rates. The goal of this paper is to find the conditions under which
the following statement will hold:%
\begin{equation}
\Xi _{N,T}\equiv \frac{1}{\sqrt{N}}\sum_{i=1}^{N}\xi _{i}\Rightarrow
\mathcal{N}(0,\Sigma _{\xi }),  \label{eq: CLT}
\end{equation}%
where
\begin{equation*}
\xi _{i}=\left(
\begin{array}{c}
\frac{1}{\sqrt{T}}\sum_{s=1}^{T}v_{s}\gamma _{i}e_{is} \\
\frac{1}{T}\sum_{s=1}^{T}\sum_{t<s}w_{st}e_{it}e_{is}%
\end{array}%
\right) ,
\end{equation*}%
and $\Sigma _{\xi }$ is an asymptotic variance matrix. Here, $e_{it}$ are weakly cross-sectionally dependent entity-specific (idiosyncratic)\footnote{%
By idiosyncratic error we mean the factor-removed part of entity-specific
variables.} errors with $\mathbb{E}\left( e_{it}\right) =0$. Errors $e_{it}$
are uncorrelated with the variables $v_{t}$ and $w_{st}$ that are common to
all units $i=1,...,N$ (more exact conditions are to appear in the next
Section). We assume $\gamma _{i},$ $i=1,...,N,$ to be non-random
entity-specific weights. Further, we want to study the circumstances when
one can also consistently estimate the asymptotic covariance -- that is,
sufficient conditions for a statement like%
\begin{equation}
\frac{1}{N}\sum_{i=1}^{N}\xi _{i}\xi _{i}^{\prime }\overset{p}{\rightarrow }%
\Sigma _{\xi }.  \label{eq: variance}
\end{equation}

As we argue below (see Examples 1--3), statements (\ref{eq: CLT}) and (\ref%
{eq: variance}) are often needed in order to conduct statistical inferences
(testing or confidence set construction) about a structural parameter, $%
\lambda ,$ which is estimated in two steps. We consider a case when in the
first step a researcher estimates a parameter $\beta _{i}$ for each
entity/unit/state $i=1,...,N$, typically via running OLS or IV time series
regressions. A typical linear estimator can be written as $\widehat{\beta }%
_{i}=\beta _{i}+\varepsilon _{i},$ where the estimation error has the
structure $\varepsilon _{i}=\big( 1+o_{p}(1)\big) \frac{1}{T}%
\sum_{t=1}^{T}v_{t}e_{it}$, with the $o_{p}(1)$ term uniformly small over
the units. In this setting, $v_{t}$ is either a regressor common to all
entities, or a common systematic part of entity-specific regressors that
have a factor structure.\footnote{%
Our setting can accommodate entity-specific regressors, say $v_{it}$, that
have a factor structure themselves. Assume that $v_{it}=a_{i}u_{t}+u_{it},$
where $u_{t}$ is a common co-movement in the regressors and $u_{it}$ is
idiosyncratic. Then
\begin{equation*}
\varepsilon _{i}=\frac{1}{T}\sum_{t=1}^{T}v_{it}e_{it}=\frac{1}{T}%
\sum_{t=1}^{T}u_{t}(a_{i}e_{it})+\frac{1}{T}\sum_{t=1}^{T}u_{it}e_{it}=\frac{%
1}{T}\sum_{t=1}^{T}v_{t}e_{it}^{\ast },
\end{equation*}%
where $v_{t}=(u_{t},1)^{\prime }$ and $e_{it}^{\ast
}=(a_{i}e_{it},u_{it}e_{it})$.}

\paragraph{Example 1.}

There is a variety of estimation approaches that can be used at the second
step. The simplest of them is weighted averaging of the first step
estimates, viz. $\widehat{\lambda }=\frac{1}{N}\sum_{i=1}^{N}\gamma _{i}%
\widehat{\beta }_{i}.$ Such an estimator is used in Sarto (2018). In order to
justify asymptotic Gaussianity of $\widehat{\lambda }$ and to make
statistical inferences about $\lambda ,$ one needs statements on the
asymptotic behavior of%
\begin{equation*}
\sqrt{\frac{T}{N}}\sum_{i=1}^{N}\gamma _{i}\varepsilon _{i}=\frac{1}{\sqrt{N}%
}\sum_{i=1}^{N}\frac{1}{\sqrt{T}}\sum_{t=1}^{T}v_{t}\gamma _{i}e_{it}.
\end{equation*}%
Note that the last expression has the structure of normalized averages stated
as the first component of $\Xi _{N,T}$ from equation (\ref{eq: CLT}). Such `linear' terms, where only the
first component of $\Xi _{N,T}$ is involved, are very common in asymptotic
derivations in factor models (e.g., Bai \& Ng 2006, 2010).

\paragraph{Example 2.}

The second estimation step may invoke a more complex estimator involving a
sample covariance between multiple first stage estimators or estimators for
multiple first stage parameters. For example, the Fama-MacBeth procedure
employs the data on excess returns to a set of portfolios $\{r_{it},\
i=1,...,N,\ t=1,...,T\}$ and time series of a risk factor $\{F_{t},\
t=1,...,T\}.$ Namely, it uses two collections of first stage parameters --
the average return on a portfolio $\beta _{i}^{(1)}=\mathbb{E}r_{it}$ via
the sample average return $\widehat{\beta }_{i}^{(1)}=\frac{1}{T}%
\sum_{t=1}^{T}r_{it},$ and the risk exposure of a portfolio $\beta
_{i}^{(2)}=\mathrm{var}(F_{t})^{-1}\mathrm{cov}(r_{it},F_{t})$ via the time
series OLS regression of $r_{it}$ on $F_{t}$ resulting in an estimate $%
\widehat{\beta }_{i}^{(2)}$. At the second stage of the Fama-MacBeth
procedure, one runs the OLS regression of the sample average return $%
\widehat{\beta }_{i}^{(1)}$ on the portfolio risk exposure estimated at the
first step $\widehat{\beta }_{i}^{(2)}$. In this case,%
\begin{equation*}
\widehat{\lambda }=\big(\sum_{i=1}^{N}\widehat{\beta }_{i}^{(2)}\widehat{\beta }%
_{i}^{(2)}\big)^{-1}\sum_{i=1}^{N}\widehat{\beta }_{i}^{(1)}\widehat{\beta }%
_{i}^{(2)},
\end{equation*}%
the second step involves two sample covariances. If one wants to derive the
asymptotic distribution of $\widehat{\lambda },$ one needs to establish the
asymptotic distribution for a properly normalized sample covariance of the
two first step estimators $\frac{1}{N}\sum_{i=1}^{N}\widehat{\beta }%
_{i}^{(1)}\widehat{\beta }_{i}^{(2)},$ where $\widehat{\beta }%
_{i}^{(j)}=\beta _{i}^{(j)}+\varepsilon _{i}^{(j)},$ with the estimation
error having the structure $\varepsilon _{i}^{(j)}=\big( 1+o_{p}(1)\big)
\frac{1}{T}\sum_{t=1}^{T}v_{t}^{(j)}e_{it}$, the term $o_{p}(1)$ being uniform in $i$.

The normalized sample covariance of the two first step estimators contains
several terms:
\begin{equation*}
\frac{1}{\sqrt{N}}\sum_{i=1}^{N}\left( \widehat{\beta }_{i}^{(1)}\widehat{%
\beta }_{i}^{(2)}-\mathbb{E}\big[\widehat{\beta }_{i}^{(1)}\widehat{\beta }%
_{i}^{(2)}\big]\right) \hspace{3in}
\end{equation*}%
\begin{equation}
=\frac{1}{\sqrt{N}}\sum_{i=1}^{N}\left( \beta _{i}^{(1)}\varepsilon
_{i}^{(2)}+\beta _{i}^{(2)}\varepsilon _{i}^{(1)}\right) +\frac{1}{\sqrt{N}}%
\sum_{i=1}^{N}\left( \varepsilon _{i}^{(1)}\varepsilon _{i}^{(2)}-\mathbb{E}%
\big[\varepsilon _{i}^{(1)}\varepsilon _{i}^{(2)}\big]\right) .  \label{eq: some eq}
\end{equation}%
The first term on the right-hand-side of equation (\ref{eq: some eq}) is
similar to a weighted average of first step estimators and has the form of
the first component of $\Xi _{N,T}$ (treating $\beta _{i}^{(j)}$ as
constants similar to constants $\gamma _{i}$). The second term in equation (%
\ref{eq: some eq}) is more complicated and calls for a Central Limit Theorem
for quadratic forms:
\begin{equation*}
\frac{T}{\sqrt{N}}\sum_{i=1}^{N}\left( \varepsilon _{i}^{(1)}\varepsilon
_{i}^{(2)}-\mathbb{E}\big[\varepsilon _{i}^{(1)}\varepsilon _{i}^{(2)}\big]\right)
\hspace{3in}
\end{equation*}%
\begin{equation*}
=\frac{1}{T\sqrt{N}}\sum_{i=1}^{N}%
\sum_{t=1}^{T}v_{t}^{(1)}v_{t}^{(2)}\big(e_{it}^{2}-\mathbb{E}[e_{it}^{2}]\big)+\frac{1%
}{\sqrt{N}}\sum_{i=1}^{N}\frac{1}{T}\sum_{s=1}^{T}%
\sum_{t<s}w_{st}e_{it}e_{is},
\end{equation*}%
where $w_{st}=v_{s}^{(1)}v_{t}^{(2)}+v_{t}^{(1)}v_{s}^{(2)}$. Here the first
term can be treated as the first component, and the second term as the
second component of $\Xi _{N,T}$.

\paragraph{Example 3.}

Anatolyev \& Mikusheva (2018) propose a split-sample estimator as an
alternative to the Fama-MacBeth procedure for factor asset pricing. There
are three sets of parameter estimates produced at the first stage: the
sample average return $\widehat{\beta }_{i}^{(1)}=\frac{1}{T}%
\sum_{t=1}^{T}r_{it}$ and two estimates of the portfolio risk exposure
computed as OLS estimates in regressions of $r_{it}$ on $F_{t}$ on different
sub-samples, say, $\widehat{\beta }_{i}^{(2)}$ and $\widehat{\beta }%
_{i}^{(3)}$. In our notation, the usage of a sub-sample is accommodated by
setting $v_{t}=0$ for those $t$ not in the currently used sub-sample. The
second step IV estimator is constructed as an IV estimator in the regression
of $\widehat{\beta }_{i}^{(1)}$ on $\widehat{\beta }_{i}^{(2)}$ using $%
\widehat{\beta }_{i}^{(3)}$ as instrument. That is,%
\begin{equation*}
\widehat{\lambda }=\big(\sum_{i=1}^{N}\widehat{\beta }_{i}^{(3)}\widehat{\beta }%
_{i}^{(1)}\big)^{-1}\sum_{i=1}^{N}\widehat{\beta }_{i}^{(3)}\widehat{\beta }%
_{i}^{(2)}.
\end{equation*}%
In order to make inferences on $\lambda ,$ one needs to obtain the
asymptotic distribution of sample covariances between different first step
estimates. Statements like (\ref{eq: CLT}) and (\ref{eq: variance}) are
instrumental to accomplish this.\bigskip

The configuration in $\Xi _{N,T}$ and a need for statements (\ref{eq: CLT})
and (\ref{eq: variance})  occur in other situations as well.

\paragraph{Example 4.}

Pesaran \& Yamagata (2018) suggest a new test for factor pricing models that
allows many portfolios to be considered simultaneously (with $N$ and $T$
both diverging to infinity). The hypothesis of interest $H_{0}:\alpha _{i}=0$
for all $i=1,...,N$, where $\alpha _{i}$ is a pricing error for the
portfolio $i$. To estimate the pricing errors the authors use OLS estimates $%
\widehat{\alpha }_{i}$. A large number of portfolios $N$ does not allow one
to establish join Gaussianity of all $\widehat{\alpha }_{i}$ or to
consistently estimate their covariance. Pesaran \& Yamagata (2018) propose
to test the hypothesis of interest using statistics based on a weighted sum
of squares of $\widehat{\alpha }_{i}$. They create a properly normalized
statistic of the form
\begin{equation*}
\sum_{i=1}^{N}\left( \frac{\widehat{\alpha }_{i}^{2}}{\sigma _{i}^{2}}%
-1\right) ,
\end{equation*}%
where $\sigma _{i}^{2}$ are variances of pricing errors. This statistic is
directly related to the sample variance of the first step estimator, and a
statement of its asymptotic Gaussianity directly follows from (\ref{eq: CLT}%
) by the same logic as stated above. Pesaran \& Yamagata (2018) develop a
CLT for quadratic forms that can be applied in this setting. They make an
assumption that the idiosyncratic components can be filtered to make them
cross-sectionally independent.\footnote{%
See Assumptions 2 and 3 in Pesaran \& Yamagata (2018).} Here we propose an
alternative version of CLT that can be applied under less restrictive
assumptions on the cross-sectional dependence of $e_{it}$'s.

\paragraph{Example 5.}

A data-rich IV environment of Bai \& Ng (2010) is another example where our
linear-quadratic CLTs can be useful. The authors consider an IV setup with
many instruments in a panel, where the number of instruments, $N$, is
potentially higher than the number of observations, $T$. The instruments are
generated by a factor model $z_{it}=\lambda _{i}^{\prime }F_{t}+e_{it},$
with $F_{t}$ and $e_{it}$ independent of the structural error $\varepsilon
_{t},$ and time-series and cross-sectional dependence in $e_{it}$ is
allowed, though restricted. Bai \& Ng (2010) consider the bias-corrected GMM
estimator that corrects for inconsistency of the baseline GMM. It is
consistent when $N/T=O(1),$ however, its asymptotic Gaussianity is
established under a more restrictive assumption when $N/T=o(1).$ The
challenge is that when $N/T=O(1),$ the asymptotic expansion for this
estimator has, in addition to a linear term, a quadratic form in the
idiosyncratic components $e_{it}$ similar to the second component of $\Xi
_{N,T}$. Thus, using statement (\ref{eq: CLT}), an asymptotic theory could
be developed for the bias-corrected GMM estimator without having to impose $%
N/T=o(1)$.\bigskip

In most of these examples, the set of idiosyncratic components $\{e_{it},$ $%
i=1,...,N,$ $t=1,...,T\}$ cannot be regarded independent and/or identically
distributed. In most realistic applications, one is usually willing to
assume that $e_{it}$ do not have a strong (detectable) factor structure, but
still allow for some correlation between different units, which would not
affect consistency. For example, it is reasonable to think that stocks
of firms in the same industry or of the same size may react to some local
shocks and be correlated, though when averaged over all stocks (and all
industries), this co-movement of returns would have no first-order impact on
estimation.

Our attempt to be agnostic with regard to possible
cross-sectional correlation among errors and to avoid explicit modeling of
its structure whenever possible comes at a cost of more restrictive time series assumptions. In many applications of interest, it is
more credible to impose independence assumptions in a time-series
direction rather than in a cross-sectional direction. For example, the
efficient market hypothesis implies mean non-predictability of excess
returns given past history, which is equivalent to a martingale difference
property for the errors. The definition of shocks in macroeconomics
similarly presumes their time-series independence. In this paper, we assume
time-series independence, which in some cases may be weakened to the
martingale difference property or stationarity with some proper mixing
condition, but we do not pursue this generalization here.

\section{CENTRAL\ LIMIT\ THEOREM}

\label{section-CLT}

In this paper we consider asymptotics as both cross-sectional and
time-series sample sizes, $N$ and $T$, increase to infinity. We allow the
data-generating process for all variables to vary with $N$ and $T$. Define $%
\mathcal{F}$ to be a $\sigma$-algebra that contains at least the $\sigma$-algebras
generated by the full set of variables $\{v_{s},s=1,...,\infty\}$ and $%
\{w_{st}, s,t=1,...,\infty\}$ for all $s $ and $t$. It may potentially also
contain other events related to common shocks and variables, as long as
Assumption \ref{ass: errors} stated below is satisfied. We treat $\gamma
_{i} $ as non-random $k_{\gamma }\times 1$ vectors.

In order to simplify the notation, in what follows we will denote $C$ to be
a positive generic constant, independent of $N$ and $T,$ which may be
different in different equations, but does not depend on or change with $N$
or $T$. We will use the following notation: for a square matrix $A$, we
denote by $\mathrm{tr}(A)$ its trace, by $\max \mathrm{ev}(A)$ -- its
maximal eigenvalue, and by $\mathrm{dg}(A)$ a diagonal matrix of the same
size with the elements from the diagonal of $A$; $\Vert \cdot \Vert $ is the
$l_{2}$ norm for a vector or the operator norm for a matrix.

\begin{ass}
\label{ass: common vars} The random $k_{v}$-vector $v_{s}$ and $k_{w}$-vector $w_{st}$ are
measurable with respect to the $\sigma$-algebra $\mathcal{F}$ for all $s,t$, and

\begin{itemize}
\item[(i)] $\frac{1}{T}\sum_{s=1}^{T}\mathbb{E}\left( v_{s}v_{s}^{\prime
}\right) \rightarrow \Omega _{v}$ and $\frac{1}{T^2}\sum_{s=1}^{T}\sum_{t<s}%
\mathbb{E}\left( w_{st}w_{st}^{\prime }\right) \rightarrow \Omega _{w}$,
where $\Omega _{v}$ and $\Omega _{w}$ are full rank matrices;

\item[(ii)] $\max_{1\leq s\leq T}\mathbb{E}\big[ \Vert v_{s}\Vert ^{4}%
\big] <C$ and $\max_{1\leq t,s\leq T}\mathbb{E}\big[ \Vert w_{st}\Vert
^{4}\big] <C$;

\item[(iii)] $\mathbb{E}\left[ \left\Vert \frac{1}{T^2}\sum_{s=1}^{T}%
\sum_{t<s}\big(w_{st}w_{st}^{\prime }-\mathbb{E}[w_{st}w_{st}^{\prime
}]\big)\right\Vert ^{2}\right] \rightarrow 0;$

\item[(iv)] $\mathbb{E}\left[ \left\Vert \frac{1}{T}%
\sum_{s=1}^{T}\big(v_{s}v_{s}^{\prime }-\mathbb{E}[v_{s}v_{s}^{\prime
}]\big)\right\Vert ^{2}\right] \rightarrow 0.$
\end{itemize}
\end{ass}

\begin{ass}
\label{ass: loadings} $\max_{1\leq i\leq N}\Vert \gamma _{i}\Vert <C$.
\end{ass}

\begin{ass}
\label{ass: errors}

\begin{itemize}
\item[(i)] Conditional on $\mathcal{F}$, the random $N$-vectors $%
e_{t}=(e_{1t},...,e_{Nt})^{\prime }$ are serially independent, and $\mathbb{E%
}(e_{t}|\mathcal{F})=0$ for all $t$;

\item[(ii)] $\max_{1\leq i\leq N,1\leq t\leq T}\mathbb{E}\left(
e_{it}^{4}\right) <C$.
\end{itemize}
\end{ass}

Assumption \ref{ass: common vars} imposes very mild restrictions on the
time-series behavior of the common (non-entity specific) variables. For
example, the part related to $v_{t}$ is trivially satisfied if a time series
equal to $v_{t}v_{t}^{\prime }$ is weakly stationary with summable
auto-covariances. Assumption \ref{ass: loadings} restricts the influence of
any one entity in the cross-sectional average and will eventually contribute
to asymptotic negligence of the cross-sectional summands needed for the CLT.
Assumption \ref{ass: errors}(i) is a restrictive assumption which imposes
discipline on the time-series structure, and the restriction $\mathbb{E}%
(e_{t}|\mathcal{F})=0$ is a form of strict exogeneity in the first step
regression. Uniform moment boundedness in Assumption \ref{ass: errors}(ii)
is traditional.

Apparently, Assumptions \ref{ass: common vars}, \ref{ass: loadings} and \ref%
{ass: errors} are insufficient to establish a central limit theorem, and we
need to put some restrictions on the cross-sectional dependence and
dependence between idiosyncratic errors and common variables. Indeed, we
will use a change of summation ordering:
\begin{equation*}
\frac{1}{\sqrt{N}}\sum_{i=1}^{N}\xi _{i}=\left(
\begin{array}{c}
\frac{1}{\sqrt{T}}\sum_{s=1}^{T}v_{s}\left( \frac{1}{\sqrt{N}}%
\sum_{i=1}^{N}\gamma _{i}e_{is}\right) \\
\frac{1}{T}\sum_{s=1}^{T}\sum_{t<s}w_{st}\left( \frac{1}{\sqrt{N}}%
\sum_{i=1}^{N}e_{it}e_{is}\right)%
\end{array}%
\right) ,
\end{equation*}%
and establish asymptotic convergence in the time-series direction. In
order to apply a CLT in the time series direction we need some sort of asymptotic
negligibility of summands with different time indexes, in particular, of
terms like $\big\{ v_{s}(\frac{1}{\sqrt{N}}\sum_{i=1}^{N}\gamma
_{i}e_{is})\big\} _{s}$ and $\big\{ w_{st}(\frac{1}{\sqrt{N}}%
\sum_{i=1}^{N}e_{it}e_{is})\big\} _{s,t}$. Our goal is to provide
low-level assumptions. There is a trade-off in how much dependence of
idiosyncratic errors across entities and how much dependence between
idiosyncratic errors and common variables can be allowed. Below we consider two
particular cases. In the first case, full independence between the $e_{it}$%
's and $\mathcal{F}$ is assumed; as a result, we can be agnostic about the
structure of cross-sectional dependence, the corresponding assumptions about
it are relatively mild. In the second case, we allow for conditional
heteroscedasticity in $e_{it}$ that can be related to some common variables
from $\mathcal{F}$ producing dependence in higher-order conditional moments.
This flexibility comes at the cost of imposing some structure on the
cross-sectional behavior of~$e_{it}$.

\subsection{Independence from common variables}

\begin{ass}
\label{ass: independence}

\begin{itemize}
\item[(i)] The errors $e_{t}=(e_{1t},...,e_{Nt})^{\prime },$ $t=1,\ldots ,T$
are independent from the $\sigma$-algebra $\mathcal{F}$ and identically
distributed across $t$;

\item[(ii)] For the $N\times N$ covariance matrix $\mathcal{E}_{N,T}=\mathbb{%
E}\left( e_{t}e_{t}^{\prime }\right) $, $\limsup_{N,T\rightarrow \infty
}\max \mathrm{ev}\left( \mathcal{E}_{N,T}\right) <\infty ,$ and $\frac{1}{N}%
\mathrm{tr}\big(\mathcal{E}_{N,T}^{2}\big)\rightarrow a<\infty$;

\item[(iii)] $\frac{1}{N}\gamma ^{\prime }\mathcal{E}_{N,T}\gamma
\rightarrow \Gamma _{\sigma }$, where $\Gamma _{\sigma }$ is full rank;

\item[(iv)] $\frac{1}{N^{2}}\sum_{i_{1}=1}^{N}\sum_{i_{2}=1}^{N}%
\sum_{i_{3}=1}^{N}\sum_{i_{4}=1}^{N}\left\vert \mathbb{E}\left(
e_{i_{1}t}e_{i_{2}t}e_{i_{3}t}e_{i_{4}t}\right) \right\vert <C.$
\end{itemize}
\end{ass}

\begin{theorem}
\label{the: gaussianity for independence} Under Assumptions \ref{ass: common
vars}, \ref{ass: loadings}, \ref{ass: errors} and \ref{ass: independence},
the central limit theorem stated in equation (\ref{eq: CLT}) holds with $\Sigma_{\xi}=\left(
                                                                                        \begin{array}{cc}
                                                                                          \Sigma_V & 0 \\
                                                                                          0 & \Sigma_W \\
                                                                                        \end{array}
                                                                                      \right)
$, where $\Sigma_V=\Gamma_\sigma\otimes\Omega_v$ and $\Sigma_W=a\Omega_w.$
\end{theorem}

Numerous papers that establish inferences in factor models commonly assume
that the set of factors is independent from the set of idiosyncratic errors,
as in Assumption \ref{ass: independence}(i), though cross-sectional
dependence of errors is allowed; see, for example, Assumption D in Bai \& Ng
(2006). We intended for the first part of Assumption \ref{ass: independence}%
(ii) to impose weak cross-sectional dependence as expressed by the
covariance matrix; in particular, it means that no strong factor structure
is left in the errors; similar assumptions appear in Onatski (2012) and Bai
\& Ng (2006). The convergence of the trace in Assumptions \ref{ass:
independence}(ii) and \ref{ass: independence}(iii) is needed for the
asymptotic covariance matrix to be properly defined.

Assumption \ref{ass: independence}(iv) is another way to restrict pervasive
dependence in multiple variables, in particular, precluding outliers to
realize in too many error terms simultaneously. For example, imagine that
the cross-sectional dependence is induced by several groups with a factor
structure, e.g., stock returns are correlated because there are
industry-specific shocks and geography-specific shocks. Imagine that there
are a finite number, say $G$, groups, indexed by $g=1,...,G$, \ which may be
overlapping, with each having independent shocks $f_{g,t}$ at time $t$.
Stock $i$ has non-zero loading $\pi _{i,g}$ only if it belongs to group $g$.
Let the set of groups, to which $i$ belongs, be denoted by $G(i)$. That is,
\begin{equation*}
e_{it}=\sum_{g\in G(i)}\pi _{i,g}f_{g,t}+\eta _{it},
\end{equation*}%
where $\eta _{it}$'s are independent both cross-sectionally and across time
and have finite fourth cumulants. Then, Assumption \ref{ass: independence}%
(iv) is essentially equivalent to the following two conditions: $\mathbb{E}%
(f_{g,t}^{4})<C$ and $\frac{1}{N}\big( \sum_{i=1}^{N}|\pi _{i,g}|\big)
^{2}<C$ for any $g=1,...,G$.  Thus,
for this example, essentially Assumption \ref{ass: independence}(iv)
imposes that the factors $f_{g,t}$ do not produce outliers too often
expressed as the moment condition and a statement about pervasiveness.

One of the important steps in the proof of Theorem \ref{the: gaussianity for
independence} verifies asymptotic negligibility of time-series summands by
checking boundedness of the fourth moments of the cross-sectional sums $%
\frac{1}{\sqrt{N}}\sum_{i=1}^{N}\gamma _{i}e_{is}$ and $\frac{1}{\sqrt{N}}%
\sum_{i=1}^{N}e_{it}e_{is}$; that imposes the main way we restrict
cross-sectional dependence. The fourth cumulant conditions are reminiscent
of those in de Jong (1987), which we follow while proving our CLT using Heyde
\& Brown (1970). There are alternative CLTs for quadratic forms such as
Rotar' (1973) that imposes weaker moment conditions on the summands but
stricter assumptions on the negligibility of coefficients and eigenvalues of
the quadratic form. In our case, following them would require imposing
stronger assumptions on the variables $w_{st},$ which we would like to avoid. Another CLT for quadratic forms for time series data can be obtained using Bhansali et al. (2007).
The book by Giraitis et al. (2012) has a chapter on this subject and allows for long memory time series as well.

\subsection{Conditional heteroscedasticity}

Assumption \ref{ass: independence}(i) of independence is much stronger than
Assumption \ref{ass: errors}(i) about exogeneity: it does not allow higher
conditional moments of $e_{it}$ to co-move with the common variables; in
particular, it imposes conditional homoscedasticity. It may be especially
problematic in financial applications where time-varying volatility is of
strong empirical relevance, and returns on many stocks display patterns of
changing volatility driven by some common variables. The assumptions below
allow for conditional heteroscedasticity.

\begin{ass}
\label{ass: heterosk} The errors $e_{it}$ have the following
weak (unobserved) factor structure:
\begin{equation*}
e_{it}=\pi _{i}^{\prime }f_{t}+\eta _{it},
\end{equation*}%
where the following assumptions hold:

\begin{itemize}
\item[(i)] The $k_{f}\times 1$ process $f_{t}$, where $k_{f}$ is fixed, is serially independent,
conditionally on $\mathcal{F}$, with $\mathbb{E}(f_{t}|\mathcal{F})=0$, $%
\mathbb{E}(f_{t}f_{t}^{\prime })=I_{k_{f}},$ $\max_{1\leq t,s\leq T}\mathbb{E%
}\big[ (\Vert v_{s}\Vert ^{4}+1)\Vert f_{t}\Vert ^{4}\big] <C,$ and $%
\max_{1\leq s,t,t^{\ast }\leq T}\mathbb{E}\big[ \Vert w_{st^{\ast }}\Vert
^{4}\Vert f_{t}\Vert ^{8}\big] <C$;

\item[(ii)] $\max \mathrm{ev}\big( \sum_{i=1}^{N}\pi _{i}\pi _{i}^{\prime
}\big) <C$ and $\frac{1}{\sqrt{N}}\sum_{i=1}^{N}\pi _{i}\gamma
_{i}^{\prime }\rightarrow \Gamma _{\pi \gamma }$;

\item[(iii)] The random variables $\eta _{it}$ are independent both
cross-sectionally and across time, independent from both $f_{s}$'s and $%
\mathcal{F}$, have mean zero and variances $\mathrm{var}(\eta _{it})=\omega
_{i}^{2}$ that are bounded from above and such that $\frac{1}{N}%
\sum_{i=1}^{N}\omega _{i}^{4}\rightarrow \omega ^{4}<\infty$, $\frac{1}{N}%
\sum_{i=1}^{N}\omega _{i}^{2}\gamma _{i}\gamma _{i}^{\prime }\rightarrow
\Gamma _{\omega }$, where $\Gamma _{\omega }$ is finite and has full rank, and $%
\max_{1\leq i\leq N,1\leq t\leq T}\mathbb{E}\left( \eta _{it}^{4}\right) <C$;

\item[(iv)] Additionally, if $\Gamma _{\pi \gamma }\neq 0$, then there
exists a matrix $\Sigma _{fv}$ such that
\begin{equation*}
\mathbb{E}\left[ \left\Vert \frac{1}{T}\sum_{s=1}^{T}(f_{s}f_{s}^{\prime
})\otimes (v_{s}v_{s}^{\prime })-\Sigma _{fv}\right\Vert ^{2}\right]
\rightarrow 0.
\end{equation*}
\end{itemize}
\end{ass}

\begin{theorem}
\label{th: CLT for HC} Under Assumptions \ref{ass: common vars}, \ref{ass:
loadings}, \ref{ass: errors} and \ref{ass: heterosk}, the statement of the
central limit theorem stated in equation (\ref{eq: CLT}) holds with $\Sigma_{\xi}=\left(
                                                                                        \begin{array}{cc}
                                                                                          \Sigma_V & 0 \\
                                                                                          0 & \Sigma_W \\
                                                                                        \end{array}
                                                                                      \right)
$, where $\Sigma_W=\omega^4\Omega_w$ and $\Sigma_V=\left(\Gamma_{\pi\gamma}^\prime \otimes I_{k_v}\right)\Sigma_{fv}\left(\Gamma_{\pi\gamma} \otimes I_{k_v}\right)+\Gamma_\omega\otimes\Omega_v$.
\end{theorem}

An interesting feature of this example is that it allows the errors to be
weakly cross-sectionally dependent to the extent that they may possess a
weak (latent) factor structure. The condition $\mathbb{E}(f_{t}f_{t}^{\prime
})=I_{k_{f}}$ is a normalization and involves no loss of generality.
Assumption \ref{ass: heterosk}(ii) forces the factors to be weak to such an
extent that the factor structure cannot be consistently detected; it implies
that the covariance matrix of idiosyncratic errors would satisfy the first
half of Assumption \ref{ass: independence}(ii). Moreover, this factor
structure may be closely related to the common variables in $\mathcal{F}$,
which causes the cross-sectional dependence among the errors $e_{it}$ to
change with the common variables and allows a very flexible form of
conditional heteroscedasticity. Indeed, the conditional cross-sectional
covariance is
\begin{equation*}
\mathbb{E}(e_{it}e_{jt}|\mathcal{F})=\pi _{i}^{\prime }\mathbb{E}%
(f_{t}f_{t}^{\prime }|\mathcal{F})\pi _{j}+\mathbb{I}_{\{i=j\}}\omega
_{i}^{2}.
\end{equation*}%
Since we do not restrict $\mathbb{E}(f_{t}f_{t}^{\prime }|\mathcal{F})$
beyond proper moment conditions, the strength of any cross-sectional
dependence as well as error variances may change stochastically depending on
realizations of the common variables.

The moment conditions in Assumption \ref{ass: heterosk}(i) help to establish
asymptotic negligibility of the time-series summands. Assumption \ref{ass:
heterosk}(iii) about $\Gamma _{\omega }$ and Assumption \ref{ass: heterosk}%
(iv) allow us to define properly the asymptotic covariance matrix.

\section{VALID\ INFERENCE}

\label{section-variance}In this Section, we first discuss estimation of
asymptotic variances for asymptotic inference when this leads to valid
inference. Then, we propose alternative tools based on the wild bootstrap to
apply in situations when asymptotic inference fails to provide
asymptotically correct inference.

\subsection{Asymptotic inference}

Statistical inferences such as confidence set construction and hypotheses
testing about the structural parameter typically require consistent
estimation of asymptotic variances of all important quantities that are
asymptotically Gaussian. The easiest to implement and thus the most
appealing from an applied perspective are those that use the same variables
and have a structure similar to the original averages, such as the statement
in equation (\ref{eq: variance}).

Notice that equation (\ref{eq: variance}) contains the cross-sectional
summation outside, and hence it treats the cross-section as nearly
uncorrelated observations, or at least it ignores the cross-sectional
correlation. A relevant analogue is the difference between the long-run
covariance and instantaneous covariance in a classical time series. However,
implementing an analogue of long-run covariance estimation here would be a
challenge since we do not have any cross-sectional stationarity or a measure
of distance between cross-sectional entities. Rather, we explore under which
conditions the convergence in (\ref{eq: variance}) holds.

Theorem \ref{th: covariance for independence} below obtains a statement for
the case when the common variables are independent from the idiosyncratic
errors, while Theorem \ref{th: covariance for HC} establishes a similar
statement for the conditionally heteroscedastic case.

\begin{theorem}
\label{th: covariance for independence} If in addition to Assumptions \ref%
{ass: common vars}, \ref{ass: loadings}, \ref{ass: errors}, \ref{ass:
independence} we also have that
\begin{equation}  \label{eq: almost diagonal}
\Vert \mathcal{E}_{N,T}-\mathrm{dg}(\mathcal{E}_{N,T})\Vert \rightarrow
0\quad \text{as }N,T\rightarrow \infty ,
\end{equation}%
then consistency statement (\ref{eq: variance}) holds.
\end{theorem}

\begin{theorem}
\label{th: covariance for HC} If in addition to Assumptions \ref{ass: common
vars}, \ref{ass: loadings}, \ref{ass: errors}, \ref{ass: heterosk} we also
have that $\Gamma _{\pi \gamma }=0$, then consistency statement (\ref{eq:
variance}) holds.
\end{theorem}

The additional assumption (\ref{eq: almost diagonal}) in Theorem \ref{th:
covariance for independence} strengthens conditions on the weakness of the
cross-sectional correlation; in particular, it requires that the covariance
matrix converges to a diagonal one. The additional assumption in Theorem \ref%
{th: covariance for HC} requires that the weights used for averaging the
cross-sectional entities are orthogonal to the loadings on the latent factor
structure, which precludes the latent factor structure (that represents the
cross-sectional dependence) from being amplified. This is a necessary
assumption for consistency of the variance estimator. Indeed, let
assumptions \ref{ass: common vars}, \ref{ass: loadings}, \ref{ass: errors}, %
\ref{ass: heterosk} hold, and consider the first component of $\xi _{i}$:
\begin{equation*}
\xi _{i}^{(1)}=\frac{1}{\sqrt{T}}\sum_{t}v_{t}\gamma _{i}e_{it}=\pi
_{i}\gamma _{i}^{\prime }\Upsilon _{T}+\tilde{\eta}_{i},
\end{equation*}%
where $\tilde{\eta}_{i}=\frac{1}{\sqrt{T}}\sum_{t}v_{t}\gamma _{i}\eta _{it}$%
, and $\Upsilon _{T}=\frac{1}{\sqrt{T}}\sum_{t=1}^{T}f_{t}v_{t}$. Note that $%
\frac{1}{\sqrt{N}}\sum_{i=1}^{N}\tilde{\eta}_{i}\Rightarrow \mathcal{N}%
(0,\sigma _{\eta }^{2})$ and $\frac{1}{N}\sum_{i=1}^{N}\tilde{\eta}_{i}^{2}%
\overset{p}{\rightarrow }\sigma _{\eta }^{2},$ as all conditions of Theorems %
\ref{th: CLT for HC} and \ref{th: covariance for HC} are satisfied by
cross-sectionally and time independent errors $\eta _{is}$. Assumption \ref%
{ass: heterosk}(iv) guarantees that $\Upsilon _{T}\Rightarrow \mathcal{N}%
(0,\Sigma _{fv})$ as $T\rightarrow \infty $, while according to Assumption %
\ref{ass: heterosk}(ii), we have $\frac{1}{\sqrt{N}}\sum_{i=1}^{N}\pi
_{i}\gamma _{i}^{\prime }\rightarrow \Gamma _{\pi \gamma }$ as $N\rightarrow
\infty $. Thus,
\begin{equation*}
\frac{1}{\sqrt{N}}\sum_{i=1}^{N}\xi _{i}^{(1)}\Rightarrow \mathcal{N}%
\big(0,\Gamma _{\pi \gamma }\Sigma _{fv}\Gamma _{\pi \gamma }^{\prime }+\sigma
_{\eta }^{2}\big),
\end{equation*}%
while $\frac{1}{N}\sum_{i=1}^{N}\big(\xi _{i}^{(1)}\big)^{2}\overset{p}{\rightarrow }%
\sigma _{\eta }^{2}$ because $\frac{1}{N}\sum_{i=1}^{N}\pi _{i}\gamma
_{i}^{\prime }\gamma _{i}\pi _{i}^{\prime }\rightarrow 0$ by Assumptions \ref%
{ass: loadings} and \ref{ass: heterosk}(ii).

\subsection{Bootstrap inference}

As one way to conduct valid inferences in settings when $\frac{1}{N}%
\sum_{i=1}^{N}\xi _{i}\xi _{i}^{\prime }$ is an inconsistent estimator of
the variance (in particular, when $\Gamma _{\pi \gamma }\neq 0$ under
Assumption \ref{ass: heterosk}), we propose the following simple wild
bootstrap procedure. In each bootstrap repetition,

\begin{enumerate}
\item[(i)] simulate independent draws of random variables $\delta _{t}\sim
\{+1,-1\},$ $t=1,...,T,$ with success probability 1/2;

\item[(ii)] compute the bootstrap analogues of errors $e_{it}^{\ast }=\delta
_{t}e_{it},$ $i=1,...,N,$ $t=1,...,T$;

\item[(iii)] compute the bootstrap analogues $\xi _{i}^{\ast }$ using the
definition of $\xi _{i},$ with $e_{it}^{\ast }$ in place of $e_{it}$ for all
$i=1,...,N,$ $t=1,...,T.$
\end{enumerate}

\noindent Then the distribution of $\Xi _{N,T}^{\ast }=\frac{1}{\sqrt{N}}%
\sum_{i=1}^{N}\xi _{i}^{\ast }$ has the same asymptotic limit as that of $%
\Xi _{N,T}$ has, and can be used for inferences. Alternatively, the
distribution of $\Xi _{N,T}^{\ast }$ normalized by the bootstrap analogue of
the variance estimate $\frac{1}{N}\sum_{i=1}^{N}\xi _{i}^{\ast }\xi
_{i}^{\ast \prime }$ can be used to approximate the distribution of $\Xi
_{N,T}$ normalized by the variance estimate $\frac{1}{N}\sum_{i=1}^{N}\xi
_{i}\xi _{i}^{\prime }.$ We call the two described bootstrap procedures
\textit{bootstrap} and \textit{bootstrap-t}. In the next Section, we
implement both variations of the wild bootstrap for the setup of Assumption %
\ref{ass: heterosk}.

The wild bootstrap works because it introduces independence in the time
direction while preserving the (unknown) cross-sectional dependence.
Specifically, we base our proof of Theorem \ref{th: CLT for HC} on the
change of order of summations in double/triple summations over $i$ and time index/indices.
For example,
\begin{equation*}
\Xi _{N,T}^{(1)}=\frac{1}{\sqrt{N}}\sum_{i=1}^{N}\xi _{i}^{(1)}=\frac{1}{%
\sqrt{T}}\sum_{t=1}^{T}\left( \frac{1}{\sqrt{N}}\sum_{i=1}^{N}v_{t}\gamma
_{i}e_{it}\right) .
\end{equation*}%
We then argue that our assumptions guarantee that the CLT with respect to
summation over $t$ is applicable. If the assumptions of the CLT hold, then $%
\Xi _{N,T}^{(1)}$ is asymptotically Gaussian with mean zero and variance
equal to the limit of $\frac{1}{T}\sum_{t=1}^T\mathbb{E}\big[\big(\frac{1}{\sqrt{N}}%
\sum_{i}v_{t}\gamma _{i}e_{it}\big)^{2}\big]$. This limit clearly depends on how much
there is cross-sectional correlation between $e_{it}$ and $e_{jt}$. Now, in
the bootstrapped samples,
\begin{equation*}
\big(\Xi _{N,T}^{(1)}\big)^{\ast }=\frac{1}{\sqrt{N}}\sum_{i=1}^{N}\big(\xi
_{i}^{(1)}\big)^{\ast }=\frac{1}{\sqrt{T}}\sum_{t=1}^{T}\left( \frac{1}{\sqrt{N}}%
\sum_{i=1}^{N}v_{t}\gamma _{i}e_{it}\right) \delta _{t}.
\end{equation*}%
Conditional on the original sample, only $\delta _{t}$'s are random, and
they are independent and have zero mean and unit variance. When $T$ is
large, this bootstrapped normalized sum satisfies the CLT and thus converges
to a zero mean Gaussian random variable with the variance equal to the limit
of $\frac{1}{T}\sum_{t=1}^{T}\big(\frac{1}{\sqrt{N}}\sum_{i=1}^{N}v_{t}\gamma
_{i}e_{it}\big)^{2}$ as $N,T\rightarrow \infty $. This limit coincides with $%
\frac{1}{T}\sum_{t=1}^{T}\mathbb{E}\big[\big(\frac{1}{\sqrt{N}}\sum_{i=1}^{N}v_{t}%
\gamma _{i}e_{it}\big)^{2}\big]$ under relatively weak assumptions, as long as the
Law of Large Numbers holds. For example, under Assumptions \ref{ass:
heterosk}, we have:
\begin{eqnarray*}
\lim \frac{1}{T}\sum_{t=1}^{T}\left( \frac{1}{\sqrt{N}}\sum_{i=1}^{N}v_{t}%
\gamma _{i}e_{it}\right) ^{2} &=&\lim \frac{1}{T}\sum_{t=1}^{T}\left( \Gamma
_{\pi \gamma }v_{t}f_{t}+\frac{1}{\sqrt{N}}\sum_{i=1}^{N}v_{t}\gamma
_{i}\eta _{it}\right) ^{2} \\
&&\overset{p}{\rightarrow }\Gamma _{\pi \gamma }\Sigma _{fv}\Gamma _{\pi
\gamma }^{\prime }+\sigma _{\eta }^{2}.
\end{eqnarray*}%
A similar argument can be made about the second component as well. Indeed,
\begin{equation*}
\big(\Xi _{N,T}^{(2)}\big)^{\ast }=\frac{1}{\sqrt{N}}\sum_{i=1}^{N}\big(\xi
_{i}^{(2)}\big)^{\ast }=\frac{1}{T}\sum_{s=1}^{T}\sum_{t<s}\left( \frac{1}{\sqrt{%
N}}\sum_{i=1}^{N}w_{st}e_{it}e_{is}\right) \delta _{t}\delta _{s}.
\end{equation*}%
For a bootstrapped statistic, conditional on the original sample, the
weights $\frac{1}{\sqrt{N}}\sum_{i=1}^{N}w_{st}e_{it}e_{is}$ are fixed,
while $\delta _{t}$'s are independent random variables, and all the
conditions of Lemma \ref{th: Heyde and brown} are satisfied. Thus, a zero
mean Gaussian limit obtains as $T\rightarrow \infty $. It is straightforward
to verify that its variance converges to the asymptotic variance of $\Xi
_{N,T}^{(2)}$.

In the simulation experiments in the following Section, we check, among
other things, that both wild bootstrap variations deliver correctly sized
tests even when the asymptotic-t tests fail to do so, and also that the
bootstrap-based tests have a non-trivial power.

\section{MONTE CARLO SIMULATIONS}

\label{section-simulations}The goals of this Section are to check finite
sample performance of an asymptotic Gaussian approximation for $\Xi _{N,T}$,
to explore when the variance estimator $\widehat{\Sigma }_{\xi }=\frac{1}{N}%
\sum_{i=1}^{N}\xi _{i}\xi _{i}^{\prime }$ allows to construct reliable
asymptotic-t inferences, and to evaluate the performance of the wild
bootstrap and bootstrap-t procedures in terms of both size and power.

\subsection{Setup}

Our setup adheres to Assumption \ref{ass: heterosk}. We generate the errors $%
e_{it}$ according to the following weak (unobserved) factor structure:
\begin{equation*}
e_{it}=\pi _{i}^{\prime }f_{t}+\eta _{it},
\end{equation*}%
where $f_{t}\sim iid$ $\mathcal{N}(0,1)$ across $t=1,...,T,$ $\gamma _{i}=1$
for all $i=1,...,N,$ and $\eta _{it}=\omega _{i}\epsilon _{\eta ,it},$ where
$\epsilon _{\eta ,it}$ are $iid$ $\mathcal{N}(0,1)$ across $i$ and $t$. The
standard deviations are set to $\omega _{i}=c_{\omega }\left( 1+\left\vert
\tau _{i}\right\vert \right) ,$ and the factor loadings are $\pi _{i}=\left(
c_{\pi }+\tau _{i}\right) /\sqrt{N},$ where $\tau _{i}\sim iid$ $\mathcal{N}%
(0,1)$ across $i=1,...,N.$ The multiplier $c_{\omega }$ is tuned so that the
average cross-sectional variance of $\eta _{it}$ is unity. The parameter $%
c_{\pi }$ indexes the degree of cross-sectional dependence as measured by
the strength of the factor structure. Specifically, as $\sum_{i=1}^{N}\pi
_{i}\pi _{i}^{\prime }\rightarrow 1+c_{\pi }^{2}$, this parameter is assumed
to be bounded for the Gaussian approximation to hold. We also have $\Gamma
_{\pi \gamma }=c_{\pi }$, hence we can expect consistency of variance
estimation only when $c_{\pi }=0$, so we will explore the distortions for
different values of $c_{\pi }$. The errors generated this way are
cross-sectionally dependent and heteroscedastic, while still satisfying
Assumption \ref{ass: heterosk}.

The common variables are generated as follows: $v_{t}=c_{fv}f_{t}+\sqrt{%
1-c_{fv}^{2}}\epsilon _{v,t},$ with $\epsilon _{v,t}\sim iid$\ $\mathcal{N}%
(0,1)$ across $t=1,...,T,$ and $w_{st}=v_{s}v_{t},$ $s,t=1,...,T.$ All the
disturbances $f_{t},$ $\tau _{i},$ $\epsilon _{\eta ,it}$ and $\epsilon
_{v,t}$ are mutually independent. The parameter $c_{fv}$ indexes the
dependence between common variables and $e_{it}$. The mean zero Assumption %
\ref{ass: errors}(i) requires $c_{fv}=0$; the non-zero values of $c_{fv}$
index deviations from the null hypothesis $\mathbb{E}(\xi _{i})=0,$ and will
be used to study the power properties of the proposed wild bootstrap. In
wild bootstrap samples, we generate the bootstrap errors by $e_{it}^{\ast
}=\delta _{t}e_{it},$ where $\delta _{t}=2\zeta _{t}-1,$ and $\zeta _{t}\sim
iid\ \mathcal{B}(\frac{1}{2})$ across $t,...,T$. The bootstrap analogues of $%
\gamma _{i},$ $v_{t}$ and $w_{st}$ are set equal to their original sample
values.

The distribution characteristics are computed from 10,000 simulations, while
the rejection rates are based on 5,000 simulation runs. In all simulations,
we set $N=T=500.$ The number of bootstrap repetitions is 600.

\subsection{Results}

Table \ref{tab: table 1} contains distributional characteristics of $\Xi
_{N,T}$. We report averages, coefficients of skewness, coefficients of
kurtosis, and right 5\% quantiles of normalized marginal distributions of
both elements of $\Xi _{N,T}.$ For the exactly normal distributions, these
values are $0$, $0$, $3$ and $1.645,$ respectively.

\begin{table}[h]
\caption{\label{tab: table 1} Characteristics of simulated finite-sample distribution of $\Xi_{N,T}$ for $N=T=500$}\medskip

\begin{centering}
\begin{tabular}{ccccccccc}
\toprule
element of $\Xi_{N,T}$$\,\rightarrow $\quad & \multicolumn{4}{c}{linear} & \multicolumn{4}{c}{
quadratic} \\
\midrule
$c_{\pi }$ $\downarrow $ & mean & skew & kurt & quant & mean & skew & kurt &
quant \\
\midrule
0.5 & $0.01$ & $-0.05$ & $3.01$ & $1.63$ & $-0.01$ & $0.25$ & $3.20$ & $1.71$
\\
1 & $0.01$ & $-0.04$ & $2.97$ & $1.64$ & $-0.01$ & $0.25$ & $3.20$ & $1.71$
\\
2 & $0.00$ & $-0.00$ & $2.95$ & $1.65$ & $0.02$ & $0.20$ & $3.11$ & $1.71$
\\
\bottomrule
\end{tabular}
\par\end{centering}
\medskip

{\footnotesize{\emph{Notes: }Based on 10,000 simulations.  The first component of  $\Xi_{N,T}$ is
labeled `linear' and second component is labeled `quadratic'. `Mean' stands for average, `skew' for skewness coefficient, `kurt' for kurtosis coefficient, and `quant' for right 5\% quantile of
simulated marginal distribution of components of  $\Xi _{N,T}$ normalized by
corresponding standard deviations. Rows with $c_{\pi }=0.5,$ $1$ and $2$
correspond to a very weak, weak and moderately strong error factor structure. }}

\end{table}

The actual distribution of the linear component of $\Xi _{N,T}$ is very
close to Gaussian, in all respects: all the moments and the right tail are
almost equal to their theoretical counterparts. The quadratic component of $%
\Xi _{N,T},$ however, albeit mean unbiased, is somewhat positively skewed
and a bit leptokurtic. The shifted right quantile confirms slight
over-dispersion. The distortions, however, do not seem to increase with the
strength of the error factor structure.

In Table \ref{tab: table 2} we document the empirical rejection rates for
tests with $10\%,$ $5\%$ and $1\%$ declared size based on the asymptotic-t,
wild bootstrap and wild bootstrap-t approaches. In the asymptotic-t approach
we create a $t$-statistic using $\widehat{\Sigma }_{\xi }$ as a variance
estimator and compare it with the symmetric standard Gaussian critical
values. We also explore the performance of two wild bootstrap procedures --
one that bootstraps $\Xi _{N,T}$ and another that bootstraps the $t$%
-statistics (referred to in Table \ref{tab: table 2} as bootstrap $\xi $ and
bootstrap $t$). In both bootstrap procedures, we compare the absolute value
of the statistic from the sample to the right quantile of the absolute value
of the bootstrapped statistic. We verify the empirical size by setting $%
c_{fv}=0$ and empirical power by setting $c_{fv}=0.1$ for relatively small
deviations from the null and $c_{fv}=0.2$ for relatively large deviations
from the null.

As expected, the size of the test based on asymptotic approximations
sometimes deviates from nominal rates by a wide margin, especially for the
linear component of $\xi _{N,T},$ the gap quickly increasing with the
strength of the error factor structure. This happens due to inconsistency of
the variance estimator $\widehat{\Sigma }_{\xi }$ and becomes more pronounced
with stronger cross-sectional dependence. What is surprising is that for the 
quadratic component of $\xi _{N,T},$ the distortions are relatively minor and 
not very sensitive to the strength of the factor structure. In contrast, 
both bootstrap procedures exhibit excellent
size control and stability thereof across the strength of the error factor
structure for both components of $\xi _{N,T}.$ In terms of power, however,
the two bootstrap statistics are approximately equally powerful for the
linear component of $\xi _{N,T},$ while there is a gap, sometimes sizable,
between power figures for its quadratic component. It seems that
bootstrapping the statistic itself is preferable.

\begin{landscape}
\begin{table}
\caption{\label{tab: table 2} Simulated rejection rates for asymptotic and wild bootstrap tests}\medskip

\begin{centering}
\begin{tabular}{ccccccccccccccccccc}
\toprule
element $\rightarrow $\quad & \multicolumn{9}{c}{linear} & \multicolumn{9}{c}{
quadratic} \\
\midrule
& \multicolumn{3}{c}{1\%} & \multicolumn{3}{c}{5\%} & \multicolumn{3}{c}{10\%
} & \multicolumn{3}{c}{1\%} & \multicolumn{3}{c}{5\%} & \multicolumn{3}{c}{
10\%} \\
\midrule
$c_{\pi }$ $\downarrow $& asy & \multicolumn{2}{c}{bootstrap} & asy & \multicolumn{2}{c}{bootstrap} & asy & \multicolumn{2}{c}{bootstrap} & asy & \multicolumn{2}{c}{bootstrap} &asy & \multicolumn{2}{c}{bootstrap} & asy & \multicolumn{2}{c}{bootstrap} \\
\midrule
 & t & $\xi$ & t & t & $\xi$ & t& t & $\xi$ & t& t & $\xi$ & t & t & $\xi$ & t & t & $\xi$ & t\\ \midrule
& \multicolumn{18}{c}{Size: $c_{fv}=0$} \\
\midrule
0.5 & 2 & 1 & 1 & 8 & 5 & 5 & 14 & 10 & 10 & 0 & 1 & 1 & 3 & 6 & 5 & 6 & 10
& 10 \\
1 & 7 & 1 & 1 & 16 & 5 & 5 & 24 & 10 & 10 & 0 & 1 & 1 & 3 & 6 & 5 & 6 & 10 & 10
\\
2 & 25 & 1 & 1 & 38 & 5 & 5 & 46 & 10 & 10 & 0 & 1 & 1 & 3 & 6 & 6 & 7 & 11
& 12 \\
\midrule
& \multicolumn{18}{c}{Power: $c_{fv}=0.1$} \\
\midrule
0.5 & 9 & 6 & 6 & 22 & 17 & 17 & 32 & 27 & 26 & 0 & 2 & 1 & 3 & 7 & 6 & 7 &
12 & 12 \\
1 & 39 & 16 & 16 & 56 & 35 & 35 & 67 & 47 & 46 & 1 & 2 & 2 & 3 & 7 & 6 & 6 &
13 & 11 \\
2 & 79 & 29 & 29 & 87 & 51 & 51 & 90 & 63 & 63 & 1 & 6 & 3 & 5 & 14 & 9 & 11
& 20 & 17 \\
\midrule
& \multicolumn{18}{c}{Power: $c_{fv}=0.2$} \\
\midrule
0.5 & 35 & 28 & 26 & 58 & 51 & 49 & 68 & 63 & 62 & 1 & 5 & 3 & 5 & 13 & 9 & 10
& 19 & 16 \\
1 & 90 & 73 & 71 & 96 & 88 & 87 & 97 & 93 & 93 & 2 & 10 & 5 & 9 & 21 & 15 &
17 & 30 & 25 \\
2 & 100 & 93 & 92 & 100 & 98 & 98 & 100 & 99 & 99 & 18 & 45 & 29 & 40 & 61 &
50 & 53 & 68 & 61 \\
\bottomrule
\end{tabular}

\end{centering}
\medskip

{\footnotesize{\emph{Notes: } The table contains actual rates, computed from 5,000
simulations for $10\%,$ $5\%$ and $1\%$ declared size
asymptotic-t (`asy t'), bootstrap (`bootstrap $\xi$') and bootstrap-t (`bootstrap t') two-sided tests for deviations of
each component of $\Xi _{N,T}$ from the zero value.
The first component of $\Xi _{N,T}$ is labeled `linear' and second component is
labeled `quadratic'. The size figures are in
panel with $c_{fv}=0,$ and power figures are in panels with $c_{fv}\neq 0.$
Rows with $c_{\pi }=0.5,$ $1$ and $2$ correspond to a very weak, weak and
moderately strong error factor structure. }}

\end{table}
\end{landscape}

\section{CONCLUDING REMARKS}

Possible directions for future research may be relaxing the error time-series independence to martingale difference structures and inventing ways to consistently estimate the asymptotic variance matrix when it is not diagonal in the limit. 
Other interesting areas involve establishing formal properties of the proposed wild bootstrap schemes, exploring the possibility of asymptotic refinements, and examining the superiority of one bootstrap scheme over the other.

\appendix
\setcounter{equation}{0}
\renewcommand{\theequation}{A.\arabic{equation}}

\section{APPENDIX: PROOFS}

\subsection{Preliminary results}

We use the following central limit theorem for a vector valued
martingale difference sequence:

\begin{lemma}
\label{th: Heyde and brown} Let the sequence $(Z_{t,T},\mathcal{F}_{t,T}),$ $%
t=1,..,T,$ be a martingale difference sequence of $r\times 1$ random vectors
with $\Sigma _{T}=\mathrm{var}\big( \sum_{t=1}^{T}Z_{t,T}\big) $. If the
following two conditions hold as $T\rightarrow \infty ,$

\begin{enumerate}
\item[(1)] $(\min \mathrm{ev}(\Sigma _{T}))^{-2}\,\sum_{t=1}^{T}\mathbb{E}%
\big[ \Vert Z_{t,T}\Vert ^{4}\big] \rightarrow 0,$

\item[(2)] $(\min \mathrm{ev}(\Sigma _{T}))^{-2}\,\mathbb{E}\big[
\big\Vert \sum_{t=1}^{T}Z_{t,T}Z_{t,T}^{\prime }-\Sigma _{T}\big\Vert ^{2}%
\big] \rightarrow 0,$
\end{enumerate}

then, as $T\rightarrow \infty $,
\begin{equation*}
\Sigma _{T}^{-1/2}\sum_{t=1}^{T}Z_{t,T}\Rightarrow \mathcal{N}(0,I_{r}).
\end{equation*}
\end{lemma}

\paragraph{Proof of Lemma \protect\ref{th: Heyde and brown}}

Indeed, the statement of Lemma \ref{th: Heyde and brown} holds if for any
non-random $r\times 1$ vector $\lambda ,$ we have $(\lambda ^{\prime }\Sigma
_{T}\lambda )^{-1/2}\sum_{t=1}^{T}\lambda ^{\prime }Z_{t,T}\Rightarrow
\mathcal{N}(0,1).$ Let us define a scalar martingale difference sequence $%
z_{t}=\lambda ^{\prime }Z_{t,T}$ with variance $\sigma _{T}^{2}=\mathrm{var}%
\big( \sum_{t=1}^{T}\lambda ^{\prime }Z_{t,T}\big) =\lambda ^{\prime
}\Sigma _{T}\lambda $. Let us check that all conditions of the central limit
theorem by Heyde \& Brown (1970) are satisfied for $\delta =1$. Indeed,
\begin{equation*}
\frac{1}{\sigma _{T}^{4}}\sum_{t=1}^{T}\mathbb{E}\left[ |z_{t}|^{4}\right] =%
\frac{1}{(\lambda ^{\prime }\Sigma _{T}\lambda )^{2}}\sum_{t=1}^{T}\mathbb{E}%
\left[ |\lambda ^{\prime }Z_{t,T}|^{4}\right] \leq \frac{1}{(\Vert \lambda
\Vert ^{2}\min \mathrm{ev}(\Sigma _{T}))^{2}}\sum_{t=1}^{T}\Vert \lambda
\Vert ^{4}\mathbb{E}\left[ \Vert Z_{t,T}\Vert ^{4}\right] \rightarrow 0,
\end{equation*}%
and%
\begin{align*}
\mathbb{E}\left[ \left\vert \frac{\sum_{t=1}^{T}z_{t}^{2}}{\sigma _{T}^{2}}%
-1\right\vert ^{2}\right] & =\mathbb{E}\left[ \left\vert \frac{%
\sum_{t=1}^{T}(\lambda ^{\prime }Z_{t,T})^{2}}{\lambda ^{\prime }\Sigma
_{T}\lambda }-1\right\vert ^{2}\right] \\
& =\frac{1}{(\lambda ^{\prime }\Sigma _{T}\lambda )^{2}}\mathbb{E}\left[
\left\vert \lambda ^{\prime }\left( \sum_{t=1}^{T}Z_{t,T}Z_{t,T}^{\prime
}-\Sigma _{T}\right) \lambda \right\vert ^{2}\right] \\
& \leq \frac{1}{(\Vert \lambda \Vert ^{2}\min \mathrm{ev}(\Sigma _{T}))^{2}}%
\Vert \lambda \Vert ^{4}\mathbb{E}\left[ \left\Vert
\sum_{t=1}^{T}Z_{t,T}Z_{t,T}^{\prime }-\Sigma _{T}\right\Vert ^{2}\right]
\rightarrow 0.
\end{align*}%
These two conditions imply that $\sigma
_{T}^{-1}\sum_{t=1}^{T}z_{t}\Rightarrow \mathcal{N}(0,1)$. This finishes the
proof. $\blacksquare $

As a preliminary result, we establish a central limit theorem for quadratic
forms. The idea of this result comes from the CLT for quadratic forms by de
Jong (1987). All random variables are implicitly indexed by the sample sizes
$T$ (or $N,T$ in the further application to factor models), which are
omitted to reduce clutter; for example, $W_{st}$ in full notation is indexed
as $W_{st,T}$ or $W_{st,N,T}$.

\begin{lemma}
\label{lem: in between lemma} Let $W_{st}=W_{st}(X_{st},e_{s},e_{t})$ be a
set of random vectors defined for all $s>t,$ where $s,t\in \{1,...,T\},$
such that $X_{st}$ is a random vector measurable with respect to the $\sigma
$-algebra $\mathcal{F}$, and all $e_{t}$ are independent from each other,
conditionally on $\mathcal{F}$. Assume that
\begin{equation}
\mathbb{E}(W_{st}|\mathcal{F},e_{t})=0\text{ and }\mathbb{E}(W_{st}|\mathcal{%
F},e_{s})=0.  \label{eq: condition on independence}
\end{equation}%
Define $W(T)=\sum_{s=1}^{T}\sum_{t<s}W_{st}$ and $\Sigma _{W,T}=\mathrm{var}%
(W(T))$. Assume the following statements hold as $T\rightarrow \infty $:

\begin{enumerate}
\item[(i)] $\Sigma_{W,T}\to \Sigma_W$, where $\Sigma_W$ is a full rank
matrix;

\item[(ii)] $T^{4}\max_{1\leq t,s\leq T}\mathbb{E}\big[ \Vert W_{st}\Vert
^{4}\big] <C$;

\item[(iii)] $\mathbb{E}\big[ \big\Vert
\sum_{s=1}^{T}\sum_{t<s}W_{st}W_{st}^{\prime }-\Sigma _{W,T}\big\Vert ^{2}%
\big] \rightarrow 0$;

\item[(iv)] $T^{4}\max_{\substack{ s_{1}\neq s_{2},t_{1}\neq t_{2}  \\ %
t_{1}<s_{1},t_{2}<s_{2}}}\,\big\vert \mathbb{E}\left(
W_{s_{1}t_{2}}^{\prime }W_{s_{2}t_{1}}W_{s_{2}t_{2}}^{\prime
}W_{s_{1}t_{1}}\right) \!\big\vert \rightarrow 0.$
\end{enumerate}

Then, as $T\rightarrow \infty $,%
\begin{equation*}
W(T)\Rightarrow \mathcal{N}(0,\Sigma _{W}).
\end{equation*}
\end{lemma}

\begin{lemma}
\label{lem: everything stacked} Let $W_{st}=W_{st}(X_{st},e_{s},e_{t})$
satisfy all conditions of Lemma \ref{lem: in between lemma}. Let $%
V_{s}=V_{s}(X_{s},e_{s})$ be a random vector defined for all $s\in
\{1,...,T\}$ such that $X_{s}$ is a random vector measurable with respect to
the $\sigma $-algebra $\mathcal{F}$, and $\mathbb{E}(V_{s}|\mathcal{F})=0.$
Define $V(T)=\sum_{s=1}^{T}V_{s}$ and $\Sigma _{V,T}=\mathrm{var}(V(T))$.
Assume the following statements hold as $T\rightarrow \infty $:

\begin{enumerate}
\item[(a)] $\Sigma_{V,T}\to \Sigma_V$, where $\Sigma_V$ is a full rank
matrix;

\item[(b)] $T\max_{1\leq s\leq T}\mathbb{E}\big[ \Vert V_{s}\Vert ^{4}%
\big] \rightarrow 0$;

\item[(c)] $\mathbb{E}\left[ \left\Vert \sum_{s=1}^{T}V_{s}V_{s}^{\prime
}-\Sigma _{V,T}\right\Vert ^{2}\right] \rightarrow 0$;

\item[(d)] $T^{3}\max_{1\leq t<\min \{s_{1},s_{2}\}\leq T}\,\left\Vert
\mathbb{E}\left( W_{s_{1}t}V_{s_{1}}^{\prime }V_{s_{2}}W_{s_{2}t}^{\prime
}\right) \right\Vert \rightarrow 0.$
\end{enumerate}

Then, as $T\rightarrow \infty $
\begin{equation*}
\left(
\begin{array}{c}
V(T) \\
W(T)%
\end{array}%
\right) \Rightarrow \mathcal{N}\left( \binom{0}{0},\left(
\begin{array}{cc}
\Sigma _{V} & 0 \\
0 & \Sigma _{W}%
\end{array}%
\right) \right) .
\end{equation*}
\end{lemma}

\paragraph{Proof of Lemma \protect\ref{lem: in between lemma}} The proof of this Lemma follows closely the proof and ideas stated in de Jong (1987).
Call $W_{st}$ \emph{clean} if
\begin{equation*}
\mathbb{E}\left( W_{s_{1}t_{1}}\otimes W_{s_{2}t_{2}}....\otimes
W_{s_{k}t_{k}}\right) =0
\end{equation*}%
when at least one index from the set $\{s_{1},t_{1},...,s_{k},t_{k}\}$ has a
value that occurs only once. The functional form of $W_{st}$ and the
condition stated in (\ref{eq: condition on independence}) guarantee that in
our case $W_{st}$ is clean. Indeed, if, for example, the index $s_{1}$
occurs only once, then
\begin{align*}
\mathbb{E}\big( W_{s_{1}t_{1}}\otimes W_{s_{2}t_{2}}\otimes ....\otimes
W_{s_{k}t_{k}}\big) & =\mathbb{E}\big[ \mathbb{E}(W_{s_{1}t_{1}}\otimes
W_{s_{2}t_{2}}\otimes ....\otimes W_{s_{k}t_{k}}|\mathcal{F}%
,e_{t_{1}},e_{s_{2}},e_{t_{2}},...,e_{t_{k}})\big] \\
& =\mathbb{E}\big[ \mathbb{E}(W_{s_{1}t_{1}}|\mathcal{F}%
,e_{t_{1}},e_{s_{2}},e_{t_{2}},...,e_{t_{k}})\otimes W_{s_{2}t_{2}}\otimes
....\otimes W_{s_{k}t_{k}}\big] \\
& =\mathbb{E}\big[ \mathbb{E}(W_{s_{1}t_{1}}|\mathcal{F},e_{t_{1}})\otimes
W_{s_{2}t_{2}}\otimes ....\otimes W_{s_{k}t_{k}}\big] =0.
\end{align*}%
Now, $W(T)=\sum_{s=1}^{T}\sum_{t<s}W_{st}=\sum_{s=1}^{T}Z_{s,T},$ where $%
Z_{s,T}=\sum_{t<s}W_{st}$. We denote by $\mathcal{F}_{s}$ the $\sigma $%
-algebra generated by $\mathcal{F}$ and $e_{t}$ for all $t<s$. Then, $%
(Z_{s,T},\mathcal{F}_{s})$ is a martingale difference sequence. Below we
check that all conditions of Lemma \ref{th: Heyde and brown} are satisfied.

Condition (i) implies that $\min \mathrm{ev}(\Sigma _{W,T})\rightarrow C>0$.
Now let us check condition (1) of Lemma \ref{th: Heyde and brown}:
\begin{align*}
\mathbb{E}\left[ \Vert Z_{s,T}\Vert ^{4}\right] & =\mathbb{E}\left[
\left\Vert \sum_{t<s}W_{st}\right\Vert ^{4}\right] \\
& =\mathbb{E}\left[ \left( \sum_{t_{1}<s}W_{st_{1}}\right) ^{\prime }\left(
\sum_{t_{2}<s}W_{st_{2}}\right) \left( \sum_{t_{3}<s}W_{st_{3}}\right)
^{\prime }\left( \sum_{t_{4}<s}W_{st_{4}}\right) \right] \\
& \leq \sum_{t<s}\mathbb{E}\left[ \Vert W_{st}\Vert ^{4}\right]
+C\sum_{t_{1}<s}\sum_{t_{2}<s,t_{2}\neq t_{1}}\mathbb{E}\left[ \Vert
W_{st_{1}}\Vert ^{2}\Vert W_{st_{2}}\Vert ^{2}\right] .
\end{align*}%
The last statement follows from the fact that $W_{st}$ is clean, and
non-zero summands are only those where either $t_{1}=t_{2}=t_{3}=t_{4}$ or
the set $\{t_{1},t_{2},t_{3},t_{4}\}$ consists of two distinct elements each
occurring twice. We also notice that $\mathbb{E}\big[ \Vert W_{st_{1}}\Vert
^{2}\Vert W_{st_{2}}\Vert ^{2}\big] \leq \frac{1}{2}\left( \mathbb{E}\big[
\Vert W_{st_{1}}\Vert ^{4}\big] +\mathbb{E}\big[ \Vert W_{st_{2}}\Vert
^{4}\big] \right) \leq \max_{1\leq t,s\leq T}\mathbb{E}\big[ \Vert
W_{st}\Vert ^{4}\big] <CT^{-4}$ due to condition (ii). Hence, $\mathbb{E}%
\big[ \Vert Z_{s,T}\Vert ^{4}\big] \leq CT^{-2}$. Thus, $\sum_{s=1}^{T}%
\mathbb{E}\big[ \Vert Z_{s,T}\Vert ^{4}\big] \leq CT^{-1}$, implying that
condition (1) of Lemma \ref{th: Heyde and brown} holds.

Now let us turn to condition (2). First, notice that%
\begin{equation*}
\Sigma _{W,T}=\mathrm{var}(W(T))=\mathrm{var}\left(
\sum_{s=1}^{T}\sum_{t<s}W_{st}\right) =\sum_{s=1}^{T}\sum_{t<s}\mathrm{var}%
(W_{st}),
\end{equation*}%
the last equality holding because $W_{st}$ is clean. Next,
\begin{align}
& \mathbb{E}\left[ \left\Vert \sum_{s=1}^{T}Z_{s,T}Z_{s,T}^{\prime }-\Sigma
_{W,T}\right\Vert _{F}^{2}\right]  \notag \\
& =\mathbb{E}\left[ \left\Vert \sum_{s=1}^{T}\left(
\sum_{t_{1}<s}W_{st_{1}}\right) \left( \sum_{t_{2}<s}W_{st_{2}}\right)
^{\prime }-\Sigma _{W,T}\right\Vert _{F}^{2}\right]  \notag \\
& =\mathbb{E}\left[ \left\Vert \sum_{s=1}^{T}\sum_{t<s}(W_{st}W_{st}^{\prime
}-\mathbb{E}[W_{st}W_{st}^{\prime }])+\sum_{s=1}^{T}\sum_{t_{1}\neq
t_{2}}W_{st_{1}}W_{st_{2}}^{\prime }\right\Vert _{F}^{2}\right]  \notag \\
& =\mathbb{E}\left[ \left\Vert \sum_{s=1}^{T}\sum_{t<s}(W_{st}W_{st}^{\prime
}-\mathbb{E}[W_{st}W_{st}^{\prime }])\right\Vert _{F}^{2}\right] +\mathbb{E}%
\left[ \left\Vert \sum_{s=1}^{T}\sum_{t_{1}\neq
t_{2}}W_{st_{1}}W_{st_{2}}^{\prime }\right\Vert _{F}^{2}\right] .
\label{eq: some 3}
\end{align}%
The last equality holds because of the clean form, as the expectation of the
Frobenius norm is equal to the trace of the sums of various products of four
terms, and any such product that contains two of the same indexes $t$ and
two different indexes $t_{1}\neq t_{2},$ has a zero expectation. Now, the
first summand in equation (\ref{eq: some 3}) converges to zero due to
condition (iii) of the Lemma. Now consider the second term in (\ref{eq: some
3}):
\begin{align*}
\mathbb{E}\left[\left\Vert \sum_{s=1}^{T}\sum_{t_{1}\neq
t_{2}<s}W_{st_{1}}W_{st_{2}}^{\prime }\right\Vert _{F}^{2}\right]&
=\sum_{s_{1}=1}^{T}\sum_{t_{1}\neq t_{2}}\sum_{s_{2}=1}^{T}\sum_{t_{3}\neq
t_{4}}\mathbb{E}\left[ \mathrm{tr}\left(
W_{s_{1}t_{1}}W_{s_{1}t_{2}}^{\prime }W_{s_{2}t_{3}}W_{s_{2}t_{4}}^{\prime
}\right) \right] \\
& =C\sum_{s_{1}=1}^{T}\sum_{s_{2}=1}^{T}\sum_{t_{1}\neq t_{2}}\mathbb{E}[%
\mathrm{tr}\left( W_{s_{1}t_{1}}W_{s_{1}t_{2}}^{\prime
}W_{s_{2}t_{1}}W_{s_{2}t_{2}}^{\prime }\right) ],
\end{align*}%
the last equality holding because $W_{st}$ is clean. The last summation can
be divided into a category when $s_{1}\neq s_{2}$, the corresponding sum
being asymptotically $o(1)$ due to condition (iv), and a category when $%
s_{1}=s_{2}$, there being at most $CT^{3}$ of such summands, each smaller
than $C\max_{1\leq t,s\leq T}\mathbb{E}\big[ \Vert W_{st}\Vert ^{4}\big]
<CT^{-4}$. Thus,
\begin{equation}
\mathbb{E}\left[ \left\Vert \sum_{s=1}^{T}\sum_{t_{1}\neq
t_{2}}W_{st_{1}}W_{st_{2}}^{\prime }\right\Vert _{F}^{2}\right] \rightarrow
0.  \label{eq some 1}
\end{equation}%
Putting statements (\ref{eq: some 3}) and (\ref{eq some 1}) together we
obtain that condition (2) of Lemma \ref{th: Heyde and brown} is satisfied.
Thus, the conclusion of Lemma \ref{lem: in between lemma} holds. $%
\blacksquare $

\paragraph{Proof of Lemma \protect\ref{lem: everything stacked}.}

Let us define $Z_{s}=(V_{s}^{\prime },\sum_{t<s}W_{st}^{\prime })^{\prime },$
and let $\mathcal{F}_{s}$ be defined as in the proof of Lemma \ref{lem: in
between lemma}. We will show that all conditions of Lemma \ref{th: Heyde and
brown} are satisfied. Notice that
\begin{equation*}
\mathbb{E}\big[V_{s}W_{st}^{\prime }\big]=\mathbb{E}\big[ \mathbb{E}%
(V_{s}W_{st}^{\prime }|\mathcal{F},e_{s})\big] =\mathbb{E}\big[V_{s}\mathbb{E}%
(W_{st}^{\prime }|\mathcal{F},e_{s})\big]=0.
\end{equation*}%
Thus,
\begin{equation*}
\Sigma _{T}=\mathrm{var}\left( \sum_{s=1}^{T}Z_{s}\right) =\left(
\begin{array}{cc}
\Sigma _{V,T} & 0 \\
0 & \Sigma _{W,T}%
\end{array}%
\right) \rightarrow \left(
\begin{array}{cc}
\Sigma _{V} & 0 \\
0 & \Sigma _{W}%
\end{array}%
\right) .
\end{equation*}%
The right-hand-side is a full rank matrix by condition (i) of Lemma \ref%
{lem: in between lemma} and condition (a) of Lemma \ref{lem: everything
stacked}. Thus, the minimal eigenvalue of $\Sigma _{T}$ is separated away
from zero for large $T$. Now,%
\begin{equation*}
\sum_{s=1}^{T}\mathbb{E}\left[ \Vert Z_{s}\Vert ^{4}\right] \leq
C\sum_{s=1}^{T}\mathbb{E}\left[ \Vert V_{s}\Vert ^{4}\right] +C\sum_{s=1}^{T}%
\mathbb{E}\left[ \left\Vert \sum_{t<s}W_{st}\right\Vert ^{4}\right] .
\end{equation*}%
The first term here is bounded by $T\max_{1\leq s\leq T}\mathbb{E}\big[
\Vert V_{s}\Vert ^{4}\big] $ which goes to zero by condition (b) of Lemma %
\ref{lem: everything stacked}, while convergence to zero of the second sum
has been already shown during the proof of Lemma \ref{lem: in between lemma}%
. Thus, condition (1) of Lemma \ref{th: Heyde and brown} holds. Next,
\begin{align*}
\mathbb{E}\left[ \left\Vert \sum_{s=1}^{T}Z_{s}Z_{s}^{\prime }-\Sigma
_{T}\right\Vert ^{2}\right] & \leq \mathbb{E}\left[ \left\Vert
\sum_{s=1}^{T}Z_{s}Z_{s}^{\prime }-\Sigma _{T}\right\Vert _{F}^{2}\right] \\
& =\mathbb{E}\left[ \left\Vert \sum_{s=1}^{T}V_{s}V_{s}^{\prime }-\Sigma
_{V,T}\right\Vert _{F}^{2}\right] +2\mathbb{E}\left[ \left\Vert
\sum_{s=1}^{T}\left( \sum_{t<s}W_{st}\right) V_{s}^{\prime }\right\Vert
_{F}^{2}\right] \\
& \quad+\mathbb{E}\left[ \left\Vert \sum_{s=1}^{T}\left( \sum_{t<s}W_{st}\right)
\left( \sum_{t<s}W_{st}\right) ^{\prime }-\Sigma _{W,T}\right\Vert _{F}^{2}%
\right] .
\end{align*}%
Here we use that the Frobenius norm of a matrix equals to the sum of squares
of all elements and can be decomposed into sums over four blocks of the
matrix. Condition (c) guarantees that
\begin{equation*}
\mathbb{E}\left[ \left\Vert \sum_{s=1}^{T}V_{s}V_{s}^{\prime }-\Sigma
_{V,T}\right\Vert _{F}^{2}\right] \leq C\mathbb{E}\left[ \left\Vert
\sum_{s=1}^{T}V_{s}V_{s}^{\prime }-\Sigma _{V,T}\right\Vert ^{2}\right]
\rightarrow 0.
\end{equation*}%
During the proof of Lemma \ref{lem: in between lemma} we show that
\begin{equation*}
\mathbb{E}\left[ \left\Vert \sum_{s=1}^{T}\left( \sum_{t<s}W_{st}\right)
\left( \sum_{t<s}W_{st}\right) ^{\prime }-\Sigma _{W,T}\right\Vert _{F}^{2}%
\right] \rightarrow 0.
\end{equation*}%
Finally,%
\begin{align*}
\mathbb{E}\left[ \left\Vert \sum_{s=1}^{T}\left( \sum_{t<s}W_{st}\right)
V_{s}^{\prime }\right\Vert _{2}^{2}\right] &
=\sum_{s_{1}=1}^{T}\sum_{t_{1}<s_{1}}\sum_{s_{2}=1}^{T}\sum_{t_{2}<s_{2}}%
\mathrm{tr}\left( \mathbb{E}\left( W_{s_{1}t_{1}}V_{s_{1}}^{\prime
}V_{s_{2}}W_{s_{2}t_{2}}^{\prime }\right) \right) \\
& =\sum_{s_{1}=1}^{T}\sum_{s_{2}=1}^{T}\sum_{t<\min \{s_{1},s_{2}\}}\mathrm{%
tr}\left( \mathbb{E}\left( W_{s_{1}t}V_{s_{1}}^{\prime
}V_{s_{2}}W_{s_{2}t}^{\prime }\right) \right) \\
& \leq CT^{3}\max_{1\leq s_{1},s_{2},t\leq T}\left\Vert \mathbb{E}\left(
W_{s_{1}t}V_{s_{1}}^{\prime }V_{s_{2}}W_{s_{2}t}^{\prime }\right)
\right\Vert \rightarrow 0.
\end{align*}%
Here we used that $\mathbb{E}\left( W_{s_{1}t_{1}}V_{s_{1}}^{\prime
}V_{s_{2}}W_{s_{2}t_{2}}^{\prime }\right) =0$ if $t_{1}\neq t_{2}$ and
condition (d) of the Lemma. To conclude, condition (2) of Lemma \ref{th:
Heyde and brown} also holds. $\blacksquare $

\begin{lemma}
\label{lemma: hadamard product} For an $N\times N$ symmetric matrix $%
A=(a_{ij})$ denote $\odot$ to be the Hadamard product. Then $\|A\odot
A\|\leq \sqrt{N}\|A\|^2. $
\end{lemma}

\paragraph{Proof.}

Using the equivalence of norms, we have%
\begin{equation*}
\Vert A\odot A\Vert \leq \Vert A\odot A\Vert _{F}=\sqrt{\sum_{1\leq i,j\leq
N}a_{ij}^{4}}\leq \sqrt{\max_{1\leq i,j\leq N}a_{ij}^{2}}\sqrt{\sum_{1\leq
i,j\leq N}a_{ij}^{2}}\leq \Vert A\Vert \Vert A\Vert _{F}\leq \sqrt{N}\Vert
A\Vert ^{2}.
\end{equation*}%
$\blacksquare $

\subsection{Proofs for Independent case}

\paragraph{Proof of Theorem \protect\ref{the: gaussianity for independence}.}

We will check that all conditions of Lemma \ref{lem: everything stacked} are
satisfied for
\begin{equation*}
W_{st}=\frac{1}{T\sqrt{N}}w_{st}\sum_{i=1}^{N}e_{it}e_{is}=\frac{1}{T}w_{st}%
\frac{e_{t}^{\prime }e_{s}}{\sqrt{N}}
\end{equation*}
and%
\begin{equation*}
V_{s}=\frac{1}{\sqrt{TN}}\sum_{i=1}^{N}\gamma _{i}e_{is}\otimes v_{s}=\frac{1%
}{\sqrt{T}}\frac{\gamma ^{\prime }e_{s}}{\sqrt{N}}\otimes v_{s}.
\end{equation*}

\noindent (i) First notice that due to Assumption \ref{ass: independence}%
(ii)
\begin{equation*}
\mathbb{E}\left[ \left( \frac{e_{t}^{\prime }e_{s}}{\sqrt{N}}\right) ^{2}%
\right] =\frac{\mathrm{tr}\big(\mathcal{E}_{N,T}^{2}\big)}{N}\rightarrow a.
\end{equation*}%
Due to the independence between the common variables and $e_{it}$ and
because $W_{st}$ is clean, we have:
\begin{equation*}
\Sigma _{W,T}=\mathrm{var}\left( \sum_{s=1}^{T}\sum_{t<s}W_{st}\right) =%
\frac{1}{T^{2}}\sum_{s=1}^{T}\sum_{t<s}\mathbb{E}(w_{st}w_{st}^{\prime })%
\mathbb{E}\left[ \left( \frac{e_{t}^{\prime }e_{s}}{\sqrt{N}}\right) ^{2}%
\right] \rightarrow a\Omega _{w},
\end{equation*}%
and the limit is a positive definite matrix.

\noindent (ii) By Assumption \ref{ass: independence}(i) and the i.i.d.
nature of $e_{t},$ we have:
\begin{equation*}
T^{4}\mathbb{E}\left[ \Vert W_{st}\Vert ^{4}\right] =\mathbb{E}\left[ \Vert
w_{st}\Vert ^{4}\right] \mathbb{E}\left[ \left( \frac{e_{t}^{\prime }e_{s}}{%
\sqrt{N}}\right) ^{4}\right] \leq \frac{C}{N^{2}}\sum_{i_{1}=1}^{N}%
\sum_{i_{2}=1}^{N}\sum_{i_{3}=1}^{N}\sum_{i_{4}=1}^{N}\mathbb{E}\big[\left(
e_{i_{1}t}e_{i_{2}t}e_{i_{3}t}e_{i_{4}t}\right) ^{2}\big]<C.
\end{equation*}%
Here we used that $|\mathbb{E}\left(
e_{i_{1}t}e_{i_{2}t}e_{i_{3}t}e_{i_{4}t}\right) |\leq \max_{1\leq i\leq
N,1\leq t\leq T}\mathbb{E}\left( e_{it}^{4}\right) <C$ and Assumption \ref%
{ass: independence}(iv).

\noindent (iii) Next,%
\begin{align*}
\sum_{s=1}^{T}\sum_{t<s}W_{st}W_{st}^{\prime }-\Sigma _{W,T}& =\frac{1}{T^{2}%
}\sum_{s=1}^{T}\sum_{t<s}w_{st}w_{st}^{\prime }\left[ \left( \frac{%
e_{s}^{\prime }e_{t}}{\sqrt{N}}\right) ^{2}-\frac{1}{N}\mathrm{tr}(\mathcal{E%
}_{N,T}^{2})\right] \\
& \quad +\frac{1}{N}\mathrm{tr}\big(\mathcal{E}_{N,T}^{2}\big)\left[ \frac{1}{T^{2}}%
\sum_{s=1}^{T}\sum_{t<s}\left( w_{st}w_{st}^{\prime }-\mathbb{E}\left(
w_{st}w_{st}^{\prime }\right) \right) \right] \\
& =A_{1}+A_{2},
\end{align*}%
hence it is enough to prove that $\mathbb{E}\big[\Vert A_{1}\Vert ^{2}\big]\rightarrow
0$ and $\mathbb{E}\big[\Vert A_{2}\Vert ^{2}\big]\rightarrow 0$. The latter is
postulated by Assumption \ref{ass: common vars}(iii). Notice that all
summands in $A_{1}$ are uncorrelated with each other due to Assumptions \ref%
{ass: errors}(i) and \ref{ass: independence}(i). Thus,
\begin{align*}
\mathbb{E}\big[ \mathrm{tr}(A_{1}A_{1}^{\prime })\big] & =\frac{1}{T^{4}}%
\sum_{s=1}^{T}\sum_{t<s}\mathbb{E}\left[ \Vert w_{st}\Vert ^{4}\right]
\mathbb{E}\left[ \left( \left( \frac{e_{s}^{\prime }e_{t}}{\sqrt{N}}\right)
^{2}-\frac{\mathrm{tr}(\mathcal{E}_{N,T}^{2})}{N}\right) ^{2}\right] \\
& \leq \frac{1}{T^{4}}\sum_{s=1}^{T}\sum_{t<s}\mathbb{E}\left[ \Vert
w_{st}\Vert ^{4}\right] \mathbb{E}\left[ \left( \frac{e_{s}^{\prime }e_{t}}{%
\sqrt{N}}\right) ^{4}\right] <\frac{C}{T^{2}}.
\end{align*}%
In the last inequality, we use the proof of statement (ii) above. This
implies that condition (iii) of Lemma \ref{lem: in between lemma} holds.

\noindent (iv) If the set $\{s_{1},s_{2},t_{1},t_{2}\}$ contains four
distinct indexes, then
\begin{align*}
T^{4}\left\vert \mathbb{E}\left( W_{s_{1}t_{2}}^{\prime
}W_{s_{2}t_{1}}W_{s_{2}t_{2}}^{\prime }W_{s_{1}t_{1}}\right) \right\vert &
\leq \mathbb{E}\left[ \Vert w_{st}\Vert ^{4}\right] \frac{\mathrm{tr}\left(
E\left( e_{s_{1}}e_{s_{1}}^{\prime }e_{t_{1}}e_{t_{1}}^{\prime
}e_{s_{2}}e_{s_{2}}^{\prime }e_{t_{2}}e_{t_{2}}^{\prime }\right) \right) }{%
N^{2}} \\
& \leq \frac{C}{N^{2}}\mathrm{tr}\big(\mathcal{E}_{N,T}^{4}\big)\leq \frac{C}{N^{2}}%
N\max \mathrm{ev}\big(\mathcal{E}_{N,T}^{4}\big)\leq \frac{C}{N}\rightarrow 0.
\end{align*}

We now move to conditions (a)-(d) of Lemma \ref{lem: everything stacked}.

\noindent (a) By Assumptions \ref{ass: independence}(iii) and \ref{ass:
common vars}(i) we have
\begin{equation*}
\Sigma _{V,T}=\left( \frac{1}{N}\gamma ^{\prime }\mathcal{E}_{N,T}\gamma
\right) \otimes \left( \frac{1}{T}\sum_{s=1}^{T}\mathbb{E}\left(
v_{s}v_{s}^{\prime }\right) \right) \rightarrow \Gamma _{\sigma }\otimes
\Omega _{v},
\end{equation*}%
and the limit is a full rank matrix.

\noindent (b) Next,
\begin{equation*}
T\mathbb{E}\left[ \Vert V_{s}\Vert ^{4}\right] =\frac{1}{T}\mathbb{E}\left[
\left\Vert \frac{1}{\sqrt{N}}\gamma ^{\prime }e_{s}\right\Vert ^{4}\right]
\mathbb{E}\left[ \Vert v_{s}\Vert ^{4}\right] ,
\end{equation*}%
where $\mathbb{E}\big[ \Vert v_{s}\Vert ^{4}\big] \leq C$ due to
Assumption \ref{ass: common vars}(ii). Assumptions \ref{ass: loadings} and %
\ref{ass: independence}(iv) imply that
\begin{align*}
\mathbb{E}\left[ \left\Vert \frac{1}{\sqrt{N}}\gamma ^{\prime
}e_{s}\right\Vert ^{4}\right] & =\frac{1}{N^{2}}\sum_{i_{1}=1}^{N}%
\sum_{i_{2}=1}^{N}\sum_{i_{3}=1}^{N}\sum_{i_{4}=1}^{N}\mathbb{E}\left(
e_{i_{1}t}e_{i_{2}t}e_{i_{3}t}e_{i_{4}t}\right) \gamma _{i_{1}}^{\prime
}\gamma _{i_{2}}\gamma _{i_{3}}^{\prime }\gamma _{i_{4}} \\
& <\max_{1\leq i\leq N}\Vert \gamma _{i}\Vert ^{4}\frac{1}{N^{2}}%
\sum_{i_{1}=1}^{N}\sum_{i_{2}=1}^{N}\sum_{i_{3}=1}^{N}\sum_{i_{4}=1}^{N}|%
\mathbb{E}\left( e_{i_{1}t}e_{i_{2}t}e_{i_{3}t}e_{i_{4}t}\right) |<C.
\end{align*}

\noindent (c) Next,
\begin{align*}
\sum_{s=1}^{T}V_{s}V_{s}^{\prime }-\Sigma _{V,T}& =\left( \frac{1}{N}\gamma
^{\prime }\mathcal{E}_{N,T}\gamma \right) \otimes \left( \frac{1}{T}%
\sum_{s=1}^{T}\big(v_{s}v_{s}^{\prime }-E\left( v_{s}v_{s}^{\prime }\right)
\big)\right) \\
& \quad +\frac{1}{T}\sum_{s=1}^{T}\left( \frac{\gamma ^{\prime
}e_{s}e_{s}^{\prime }\gamma }{N}-\frac{\gamma ^{\prime }\mathcal{E}%
_{N,T}\gamma }{N}\right) \otimes \left( v_{s}v_{s}^{\prime }\right) \\
& =A_{1}+A_{2}.
\end{align*}%
Notice that $A_{1}$ and $A_{2}$ are uncorrelated, hence
\begin{equation*}
\mathbb{E}\left[ \left\Vert \sum_{s=1}^{T}V_{s}V_{s}^{\prime }-\Sigma
_{V,T}\right\Vert _{F}^{2}\right] =\mathrm{tr}\left( \mathbb{E}%
(A_{1}^{\prime }A_{1})\right) +\mathrm{tr}\left( \mathbb{E}(A_{2}^{\prime
}A_{2})\right) .
\end{equation*}%
Assumption \ref{ass: common vars}(iv) guarantees the convergence of the
first term. Notice that the summands in $A_{2}$ are uncorrelated due to time
independence of errors, hence
\begin{align*}
\mathrm{tr}\left( \mathbb{E}(A_{2}^{\prime }A_{2})\right) & =\frac{1}{T^{2}}%
\sum_{s=1}^{T}\mathbb{E}\left[ \left\vert \frac{\gamma ^{\prime
}e_{s}e_{s}^{\prime }\gamma }{N}-\frac{1}{N}\gamma ^{\prime }\mathcal{E}%
_{N,T}\gamma \right\vert ^{2}\right] \mathbb{E}\left[ \Vert v_{s}\Vert ^{4}%
\right] \\
& \leq \frac{C}{T}\mathbb{E}\left[ \left\vert \frac{\gamma ^{\prime
}e_{s}e_{s}^{\prime }\gamma }{N}-\frac{1}{N}\gamma ^{\prime }\mathcal{E}%
_{N,T}\gamma \right\vert ^{2}\right] .
\end{align*}%
Given the bounds on the fourth moment of $N^{-1/2}\gamma ^{\prime }e_{s}$
derived in the proof of part (b) we get that condition (c) holds.

\noindent (d) By Assumption \ref{ass: independence}(i) we have that
\begin{equation*}
T^{3}\left\Vert \mathbb{E}\left( W_{s_{1}t}V_{s_{1}}^{\prime
}V_{s_{2}}W_{s_{2}t}^{\prime }\right) \right\Vert =\left\Vert \mathbb{E}%
\left( w_{s_{1}t}v_{s_{1}}^{\prime }v_{s_{2}}w_{s_{2}t}^{\prime }\right)
\mathbb{E}\left( \frac{e_{s_{1}}^{\prime }\gamma }{\sqrt{N}}\frac{\gamma
^{\prime }e_{s_{2}}}{\sqrt{N}}\frac{e_{s_{1}}^{\prime }e_{t}}{\sqrt{N}}\frac{%
e_{s_{2}}^{\prime }e_{t}}{\sqrt{N}}\right) \right\Vert .
\end{equation*}%
Using that scalars can be reshuffled to make two same-index $e_{t}$ 
stand back to back and employing time series independence of errors, we
obtain that%
\begin{align*}
\left\vert \mathbb{E}\left( \frac{e_{s_{1}}^{\prime }\gamma }{\sqrt{N}}\frac{%
\gamma ^{\prime }e_{s_{2}}}{\sqrt{N}}\frac{e_{s_{1}}^{\prime }e_{t}}{\sqrt{N}%
}\frac{e_{s_{2}}^{\prime }e_{t}}{\sqrt{N}}\right) \right\vert & =\frac{1}{%
N^{2}}\left\vert \mathrm{tr}\big(\gamma \gamma ^{\prime }\mathbb{E}%
(e_{s_{2}}e_{s_{2}}^{\prime })\mathbb{E}(e_{t}e_{t}^{\prime })\mathbb{E}%
(e_{s_{1}}e_{s_{1}}^{\prime })\big)\right\vert \\
& \leq \frac{1}{N^{2}}\mathrm{tr}(\gamma \gamma ^{\prime })\max \mathrm{ev}\big(%
\mathcal{E}_{N,T}^{3}\big)\leq \frac{C}{N}.
\end{align*}%
Here we use Assumption \ref{ass: loadings} to get $N^{-1}\mathrm{tr}%
(\gamma \gamma ^{\prime })<C$ and Assumption \ref{ass: independence}(ii).
Given Assumption \ref{ass: common vars}(ii) we obtain that
\begin{equation*}
T^{3}\max_{1\leq t<\min \{s_{1},s_{2}\}\leq T}\left\Vert \mathbb{E}\left(
W_{s_{1}t}V_{s_{1}}^{\prime }V_{s_{2}}W_{s_{2}t}^{\prime }\right)
\right\Vert \leq \frac{C}{N}\rightarrow 0.
\end{equation*}%
Thus, condition (d) of Lemma \ref{lem: everything stacked} is satisfied.
This concludes the proof of Theorem \ref{the: gaussianity for independence}.
$\blacksquare $

\paragraph{Proof of Theorem \protect\ref{th: covariance for independence}.}

We will prove the following three statements for
\begin{equation*}
\xi _{V,i}=\frac{1}{\sqrt{T}}\sum_{s=1}^{T}\gamma _{i}e_{is}\otimes v_{s}
\end{equation*}%
and
\begin{equation*}
\xi _{W,i}=\frac{1}{T}\sum_{s=1}^{T}\sum_{t<s}w_{st}e_{it}e_{is}\text{:}
\end{equation*}

\begin{itemize}
\item[(i)] $N^{-1}\sum_{i=1}^{N}\xi _{V,i}\xi _{V,i}^{\prime }\overset{p}{%
\rightarrow }\Sigma _{V}$;

\item[(ii)] $N^{-1}\sum_{i=1}^{N}\xi _{W,i}\xi _{W,i}^{\prime }\overset{p}{%
\rightarrow }\Sigma _{W}$;

\item[(iii)] $N^{-1}\sum_{i=1}^{N}\xi _{V,i}\xi _{W,i}^{\prime }\overset{p}{%
\rightarrow }0$.
\end{itemize}

Let us start with statement (i). Denote by $\sigma _{i}^{2}$ the diagonal
and by$\ \sigma _{ij}$ the off-diagonal elements of matrix $\mathcal{E}%
_{N,T} $. Notice that the additional assumption of Theorem \ref{th:
covariance for independence} implies that
\begin{equation*}
\Gamma _{\sigma }=\lim \frac{\gamma ^{\prime }\mathcal{E}_{N,T}\gamma }{N}%
=\lim \frac{1}{N}\sum_{i=1}^{N}\gamma _{i}\gamma _{i}^{\prime }\sigma
_{i}^{2}.
\end{equation*}%
Let us define $\widetilde{\Sigma }_{V,T}=\big( N^{-1}\sum_{i=1}^{N}\gamma
_{i}\gamma _{i}^{\prime }\sigma _{i}^{2}\big) \big( T^{-1}\sum_{s=1}^{T}%
\mathbb{E}\left( v_{s}v_{s}^{\prime }\right) \big) $, and notice that $%
\widetilde{\Sigma }_{V,T}\rightarrow \Sigma _{V}$. Thus,
\begin{align*}
\frac{1}{N}\sum_{i=1}^{N}\xi _{V,i}\xi _{V,i}^{\prime }-\widetilde{\Sigma }%
_{V,T}& =\frac{1}{NT}\sum_{i=1}^{N}\sum_{t=1}^{T}\sum_{s=1}^{T}\left( \left(
\gamma _{i}\gamma _{i}^{\prime }e_{is}e_{it}\right) \otimes
(v_{s}v_{t}^{\prime })-\mathbb{I}\{s=t\}\sigma _{i}^{2}\left( \gamma
_{i}\gamma _{i}^{\prime }\right) \otimes \mathbb{E}(v_{t}v_{t}^{\prime
})\right) \\
& =\frac{1}{NT}\sum_{i=1}^{N}\sum_{t=1}^{T}(e_{it}^{2}-\sigma
_{i}^{2})\left( \gamma _{i}\gamma _{i}^{\prime }\right) \otimes
(v_{t}v_{t}^{\prime }) \\
& \quad +\frac{1}{NT}\sum_{i=1}^{N}\sum_{t=1}^{T}\sum_{s\neq t}\left( \gamma
_{i}\gamma _{i}^{\prime }e_{is}e_{it}\right) \otimes (v_{s}v_{t}^{\prime })
\\
& \quad +\frac{1}{NT}\sum_{i=1}^{N}\sum_{t=1}^{T}\left( \gamma _{i}\gamma
_{i}^{\prime }\sigma _{i}^{2}\right) \otimes \big(v_{t}v_{t}^{\prime }-\mathbb{E}%
\left( v_{t}v_{t}^{\prime }\right)\!\big) \\
& =A_{1}+A_{2}+A_{3}.
\end{align*}%
Notice that the three terms are uncorrelated, so it is enough to prove that $%
\mathrm{tr}\left( \mathbb{E}(A_{j}A_{j}^{\prime })\right) \rightarrow 0$ for
$j=1,2,3.$  Indeed, if the expectation of the Frobenius norm of a matrix converges to zero, this implies that each entry converges to zero as well. First,
\begin{align*}
\mathrm{tr}\left( \mathbb{E}(A_{1}A_{1}^{\prime })\right) & =\mathrm{tr}%
\left( \mathbb{E}\left[ \frac{1}{N^{2}T^{2}}\sum_{i,j=1}^{N}\sum_{t=1}^{T}%
\sum_{s=1}^{T}\left( \gamma _{i}\gamma _{i}^{\prime }\gamma _{j}\gamma
_{j}^{\prime }(e_{it}^{2}-\sigma _{i}^{2})(e_{js}^{2}-\sigma
_{i}^{2})\right) \otimes (v_{t}v_{t}^{\prime }v_{s}v_{s}^{\prime })\right]
\right) \\
& =\frac{1}{N^{2}T^{2}}\sum_{i,j=1}^{N}\sum_{t=1}^{T}\mathrm{tr}\left(
\gamma _{i}\gamma _{i}^{\prime }\gamma _{j}\gamma _{j}^{\prime }\mathrm{cov}%
(e_{it}^{2},e_{jt}^{2})\right) \mathrm{tr}\big( \mathbb{E}%
(v_{t}v_{t}^{\prime }v_{t}v_{t}^{\prime })\big) \\
& \leq \frac{1}{T^{2}}\sum_{t=1}^{T}\max_{1\leq i\leq N}\Vert \gamma
_{i}\Vert ^{4}\max_{1\leq i\leq N}\mathbb{E}\left[ (e_{it}^{2}-\sigma
_{i}^{2})^{2}\right] \mathbb{E}\big[ \Vert v_{t}\Vert ^{4}\big] \leq
\frac{C}{T}.
\end{align*}%
Here we used that $e_{it}$'s are independent from each other for different $%
t $ by Assumption \ref{ass: errors}(i), which forces $s=t$. The last
inequality uses Assumptions \ref{ass: common vars}(ii), \ref{ass: errors}%
(ii) and \ref{ass: loadings}.

Consider the term $A_{2}$ and notice that any two summands in the
two-directional sum (over $t$ and over $s$) are uncorrelated due to time
series independence of $e_{t}$'s and all summands are mean zero. Thus,%
\begin{align*}
\mathrm{tr}\left( \mathbb{E}(A_{2}A_{2}^{\prime })\right) & =\frac{1}{%
N^{2}T^{2}}\sum_{t=1}^{T}\sum_{s\neq t}\sum_{i,j=1}^{N}\mathrm{tr}\left(
\mathbb{E}(\gamma _{i}\gamma _{i}^{\prime }\gamma _{j}\gamma _{j}^{\prime
}e_{it}e_{is}e_{jt}e_{js})\otimes \mathbb{E}(v_{s}v_{t}^{\prime
}v_{t}v_{s}^{\prime })\right) \\
& =\frac{1}{N^{2}}\sum_{i,j=1}^{N}\mathrm{tr}(\gamma _{i}\gamma _{i}^{\prime
}\gamma _{j}\gamma _{j}^{\prime }\sigma _{ij}^{2})\frac{1}{T^{2}}%
\sum_{t=1}^{T}\sum_{s\neq t}\mathrm{tr}\big( \mathbb{E}\left(
v_{s}v_{t}^{\prime }v_{t}v_{s}^{\prime }\right)\! \big) .
\end{align*}%
We notice that $T^{-2}\sum_{t=1}^{T}\sum_{s\neq t}\mathrm{tr}\big( \mathbb{E%
}\left( v_{s}v_{t}^{\prime }v_{t}v_{s}^{\prime }\right)\!\big) \leq \mathbb{%
E}\big[ \Vert v_{t}\Vert ^{4}\big] <C$ due to Assumption \ref{ass: common
vars}(ii). Denote $r,r^{\ast }$ to be indexes that go over $1,...,k_{\gamma }$.
For any fixed value of $r,r^{\ast }$ denote $B^{(r,r^{\ast
})}=\big((\gamma _{i}\gamma _{i}^{\prime })_{r,r^{\ast }}\big)_{i=1}^{N}$, an $%
N\times 1$ vector. Then,%
\begin{align*}
\frac{1}{N^{2}}\sum_{i,j=1}^{N}\mathrm{tr}(\gamma _{i}\gamma _{i}^{\prime
}\gamma _{j}\gamma _{j}^{\prime }\sigma _{ij}^{2})& =\frac{1}{N^{2}}%
\sum_{i,j=1}^{N}\sum_{r,r^{\ast }}(\gamma _{i}\gamma _{i}^{\prime
})_{r,r^{\ast }}(\gamma _{j}\gamma _{j}^{\prime })_{r,r^{\ast }}\sigma
_{ij}^{2} \\
& =\sum_{r,r^{\ast }}\frac{1}{N^{2}}\sum_{i=1}^{N}(\gamma _{i}\gamma
_{i}^{\prime })_{r,r^{\ast }}(\gamma _{i}\gamma _{i}^{\prime })_{r,r^{\ast
}}\sigma _{i}^{4} \\
& \quad +\sum_{r,r^{\ast }}\frac{1}{N^{2}}B^{(r,r^{\ast })\prime }\big[ (%
\mathcal{E}_{N,T}-\mathrm{dg}(\mathcal{E}_{N,T}))\odot (\mathcal{E}_{N,T}-%
\mathrm{dg}(\mathcal{E}_{N,T}))\big] B^{(r,r^{\ast })} \\
& \leq k_{\gamma }^{2}\max_{1\leq i\leq N}\Vert \gamma
_{i}\Vert ^{4}\left( \frac{1}{N^{2}}\sum_{i=1}^{N}\sigma _{i}^{4}+\frac{%
\sqrt{N}\Vert \mathcal{E}_{N,T}-\mathrm{dg}(\mathcal{E}_{N,T})\Vert }{N^{2}}%
\right) \\
& \leq \frac{C}{\sqrt{N}}\rightarrow 0,
\end{align*}%
where in the second to last inequality we used Lemma \ref{lemma: hadamard
product} and the last inequality is due to Assumptions \ref{ass: loadings}, %
\ref{ass: independence}(ii) and the additional assumption stated in Theorem %
\ref{th: covariance for independence}. This shows that $\mathrm{tr}\left(
\mathbb{E}(A_{2}A_{2}^{\prime })\right) \rightarrow 0.$

Finally, $\mathrm{tr}\left( \mathbb{E}(A_{3}A_{3}^{\prime })\right)
\rightarrow 0$ due to Assumption \ref{ass: common vars}(iv). This ends the
proof of statement (i).

Let us turn to statement (ii):
\begin{align*}
\frac{1}{N}\sum_{i=1}^{N}\xi _{W,i}\xi _{W,i}^{\prime }-\Sigma _{W,T}& =%
\frac{1}{T^{2}N}\sum_{i=1}^{N}\sum_{s=1}^{T}\sum_{t<s}\left(
e_{it}^{2}e_{is}^{2}-\sigma _{i}^{4}\right) w_{st}w_{st}^{\prime } \\
& \quad +\frac{1}{T^{2}N}\sum_{i=1}^{N}\sum_{s_{1}=1}^{T}\sum_{t_{1}<s_{1}}%
\sum_{s_{2}=1}^{T}\sum_{\substack{ t_{2}<s_{2},  \\ \{s_{1},t_{1}\}\neq
\{s_{2},t_{2}\}}}w_{s_{1}t_{1}}w_{s_{2}t_{2}}^{\prime
}e_{it_{1}}e_{is_{2}}e_{it_{2}}e_{is_{1}} \\
& \quad +\frac{1}{N}\sum_{i=1}^{N}\sigma _{i}^{4}\frac{1}{T^{2}}%
\sum_{s=1}^{T}\sum_{t<s}\big( w_{st}w_{st}^{\prime }-\mathbb{E}%
(w_{st}w_{st}^{\prime })\big) \\
& =A_{1}+A_{2}+A_{3}.
\end{align*}%
Again, $A_{1},A_{2}$ and $A_{3}$ are uncorrelated with each other. Thus, we
can deal with each one of them separately. We show that the expectation of the Frobenius norm of each matrix converges to zero, this implies that each entry converges to zero as well.

Let us start with
\begin{equation*}
\mathrm{tr}\left( \mathbb{E}(A_{1}A_{1}^{\prime })\right) =\frac{1}{%
T^{4}N^{2}}\sum_{i,j=1}^{N}\sum_{s,s^{\ast }=1}^{T}\sum_{t<s,t^{\ast
}<s^{\ast }}\mathrm{tr}\big( \mathbb{E}(w_{st}w_{st}^{\prime }w_{s^{\ast
}t^{\ast }}w_{s^{\ast }t^{\ast }}^{\prime })\big) \mathbb{E}%
(b_{i,t,s}b_{j,t^{\ast },s^{\ast }}),
\end{equation*}%
where
\begin{equation*}
b_{i,t,s}=e_{it}^{2}e_{is}^{2}-\sigma _{i}^{4}=(e_{it}^{2}-\sigma
_{i}^{2})(e_{is}^{2}-\sigma _{i}^{2})+\sigma _{i}^{2}(e_{is}^{2}-\sigma
_{i}^{2})+\sigma _{i}^{2}(e_{it}^{2}-\sigma _{i}^{2}).
\end{equation*}%
Notice that $\mathbb{E}(b_{i,t,s}b_{j,t^{\ast },s^{\ast }})\neq 0$ only if
at least one of the indexes from the set $\{t,t^{\ast },s,s^{\ast }\}$
appears twice. Thus, the summation over time index is three-dimensional and
there are at most $CT^{3}N^{2}$ non-zero summands in $\mathrm{tr}\left(
\mathbb{E}(A_{1}A_{1}^{\prime })\right) $. Let us bound every summand from
above. Notice that since $t<s$ and $t^{\ast }<s^{\ast },$ all indexes in the
set $\{t,t^{\ast },s,s^{\ast }\}$ can appear at most twice; also errors with
different time indexes are independent from each other, so the largest
moment of the error term we will have is the fourth. To sum up, each
non-zero summand is bounded above by $T^{-4}N^{-2}C\max_{1\leq t,s\leq T}%
\mathbb{E}\big[ \Vert w_{st}\Vert ^{4}\big] \max_{1\leq i\leq N,1\leq
t\leq T}\mathbb{E}\big[\left( e_{it}^{4}\right) ^{2}\big]$, thus $\mathrm{tr}\left(
\mathbb{E}(A_{1}A_{1}^{\prime })\right) \leq C/T\rightarrow 0.$

The term $\mathrm{tr}\left( \mathbb{E}(A_{2}A_{2}^{\prime })\right) $
includes summation over eight time indexes but most of the summands are
zeros. The non-zero terms place at least four restrictions on the time
indexes. We note that the non-trivial part of the sum in $\mathrm{tr}\left(
\mathbb{E}(A_{2}A_{2}^{\prime })\right) $ includes summation over $%
i,j=1,...,N$ and over time indexes $\{s_{1},s_{1}^{\ast },t_{1},t_{1}^{\ast
},s_{2},s_{2}^{\ast },t_{2},t_{2}^{\ast }\}$, where in the last set any
distinct index appears at least twice. The summands are
\begin{equation*}
\frac{1}{T^{4}N^{2}}\mathbb{E}\big( w_{s_{1}t_{1}}w_{s_{1}^{\ast
}t_{1}^{\ast }}^{\prime }w_{s_{2}^{\ast }t_{2}^{\ast }}^{\prime
}w_{s_{2}t_{2}}\big) \mathbb{E}\left( e_{it_{1}}e_{is_{1}}e_{it_{1}^{\ast
}}e_{is_{1}^{\ast }}e_{jt_{2}}e_{js_{2}}e_{jt_{2}^{\ast }}e_{js_{2}^{\ast
}}\right) .
\end{equation*}%
Notice also that due to restrictions that $t$'s are strictly smaller than
their corresponding $s$'s, each time index can appear at most four times,
hence we get at most fourth power of each error term.

First, consider the case when the set $\{s_{1},s_{1}^{\ast
},t_{1},t_{1}^{\ast },s_{2},s_{2}^{\ast },t_{2},t_{2}^{\ast }\}$ contains at
most three distinct indexes (this makes the summation over time
three-dimensional). We can show that each summand is bounded by $%
T^{-4}N^{-2}\max_{1\leq t,s\leq T}\mathbb{E}\big[ \Vert w_{ts}\Vert ^{4}%
\big] \max_{1\leq i\leq N,1\leq t\leq T}\mathbb{E}\big( e_{it}^{4}\big)^2
 \linebreak[0]\leq C/\left( T^{4}N^{2}\right) $ in absolute value, and as there are at
most $N^{2}T^{3}$ of them (two-dimensional cross-sectional and
three-dimensional over time summations), the sum of such terms will go to
zero.

Finally, we consider the case when the set $\{s_{1},s_{1}^{\ast
},t_{1},t_{1}^{\ast },s_{2},s_{2}^{\ast },t_{2},t_{2}^{\ast }\}$ contains
four distinct indexes. Then each summand of this type is bounded in absolute
value by
\begin{equation*}
C\frac{|\sigma _{ij}|^{a}(\sigma _{i}^2)^{b}(\sigma _{j}^2)^{c}}{T^{4}N^{2}}%
\max_{1\leq s,t\leq T}\mathbb{E}\left[ \left\Vert w_{st}\right\Vert ^{4}%
\right] ,
\end{equation*}%
where $a+b+c=4,$ and the values of $a,$ $b$ and $c$ depend on which indices
coincide with which; however, due to the conditions $\{s_{1},t_{1}\}\neq
\{s_{2},t_{2}\}$ and $t_{1}<s_{1},t_{2}<s_{2},$ we know that the set $%
\{s_{1},s_{2},t_{1},t_{2}\}$ contains at least three distinct indexes. Thus,
$c$ and $b$ are either 0 or 1 each, and $a\geq 2$. Hence, due to Assumption %
\ref{ass: common vars}(ii), the corresponding sum is bounded above by
\begin{align}
\frac{C}{N^{2}}\sum_{i=1}^{N}\sum_{j=1}^{N}|\sigma _{ij}|^{a}(\sigma
_{i}^2)^{b}(\sigma _{j}^2)^{c}& \leq \frac{C}{N^{2}}\max_{1\leq i\leq
N}\sigma _{i}^{4}\sum_{i=1}^{N}\sum_{j=1}^{N}\sigma _{ij}^{2}  \notag \\
& =\frac{C}{N^{2}}\sum_{i=1}^{N}\sigma _{i}^{4}+\frac{C}{N^{2}}%
\sum_{i=1}^{N}\sum_{i\neq j}^{N}\sigma _{ij}^{2}
\label{eq: prove of frobenius norm} \\
& \leq \frac{C}{N}  \max_{1\leq i\leq N}\sigma _{i}^{4}
 +\frac{C}{N^{2}}\Vert \mathcal{E}_{N,T}-\mathrm{dg}(%
\mathcal{E}_{N,T})\Vert _{F}^{2}
 \leq \frac{C}{N}.  \notag
\end{align}%
In the first inequality, we use $|\sigma _{ij}|\leq \sigma _{i}\sigma _{j}$.
In the second inequality, we use the definition of the Frobenius norm.
In the last inequality, we use that for any symmetric matrix $A,$ we have $\Vert
A\Vert _{F}^{2}\leq N\Vert A\Vert ^{2}$ and assumption stated in Theorem \ref%
{th: covariance for independence}. Thus, $\mathrm{tr}\left( \mathbb{E}%
(A_{2}A_{2}^{\prime })\right) \rightarrow 0.$

Next, Assumption \ref{ass: common vars}(iii) implies the convergence of $%
A_{3}$. This finishes the proof of (ii).

Finally, we need to prove statement (iii) that
\begin{equation*}
\frac{1}{NT^{3/2}}\sum_{i=1}^{N}\sum_{s=1}^{T}\sum_{t<s}\sum_{s^{\ast
}=1}^{T}(\gamma _{i}\otimes v_{s^{\ast }})w_{st}^{\prime }e_{is^{\ast
}}e_{it}e_{is}\rightarrow ^{p}0.
\end{equation*}%
As before, we look at the expectation of the square of the sum above, which
involves six-dimensional summation over time indexes and two-dimensional
summation over cross-section (over $i,j$) and is normalized by $N^{-2}T^{-3}$%
. Due to time-series independence of $e_{it}$, the six-dimensional summation
over time indexes has mostly zeros and can be reduced to three-dimensional
summation over time indexes as the set $\{s_{1},t_{1},s_{1}^{\ast
},s_{2},t_{2},s_{2}^{\ast }\}$ should have any distinct index to appear at
least twice.

First, consider only those terms for which the set $\{s_{1},t_{1},s_{1}^{%
\ast },s_{2},t_{2},s_{2}^{\ast }\}$ contains at most two distinct indexes;
there are at most $N^{2}T^{2}$ of such terms. Since $t_{1}<s_{1}$ and $%
t_{2}<s_{2},$ each time index can appear at most four times; thus, the
highest power of each individual shock can be the fourth. As a result, each
summand is bounded above by $N^{-2}T^{-3}\max_{1\leq i\leq N}\Vert \gamma
_{i}\Vert ^{2}\max_{1\leq t\leq T}\mathbb{E}\big[ \Vert v_{t}\Vert ^{2}%
\big] \max_{1\leq t,s\leq T}\mathbb{E}\big[ \Vert w_{st}\Vert ^{2}\big]
\max_{1\leq i\leq N,1\leq t\leq T}\mathbb{E}\big( e_{it}^{4}\big)^{3/2}.$
Given Assumptions \ref{ass: common vars}(ii) and \ref{ass: errors}(ii),
the sum of these terms is bounded above by $C/T$.

Finally, consider only those terms for which the set $\{s_{1},t_{1},s_{1}^{%
\ast },s_{2},t_{2},s_{2}^{\ast }\}$ contains exactly three distinct indexes.
The summation over these indexes is equal to
\begin{equation*}
\mathrm{tr}\left( \frac{1}{N^{2}}\sum_{i,j=1}^{N}\gamma _{i}\gamma
_{j}^{\prime }\big(C_{1}\sigma _{ij}\sigma _{i}^{2}\sigma _{j}^{2}+C_{2}\sigma
_{ij}^{3}\big)\right) .
\end{equation*}%
The term $\sigma _{ij}^{3}$ appears when $\{s_{1},t_{1},s_{1}^{\ast
}\}=\{s_{2},t_{2},s_{2}^{\ast }\}$, while $\sigma _{ij}\sigma _{i}^{2}\sigma
_{j}^{2}$ arises when the sets $\{s_{1},t_{1},s_{1}^{\ast }\}$ and $%
\{s_{2},t_{2},s_{2}^{\ast }\}$ have two coinciding indexes each. Therefore,
\begin{align*}
\mathrm{tr}\left( \frac{1}{N^{2}}\sum_{i,j=1}^{N}\gamma _{i}\gamma
_{j}^{\prime }\sigma _{ij}\sigma _{i}^{2}\sigma _{j}^{2}\right) & =\mathrm{tr%
}\left( \frac{1}{N^{2}}\sum_{i=1}^{N}\gamma _{i}\gamma _{i}^{\prime }\sigma
_{i}^{6}\right) +\mathrm{tr}\left( \frac{1}{N^{2}}\sum_{i\neq j}(\gamma
_{i}\sigma _{i}^{2})(\gamma _{j}^{\prime }\sigma _{j}^{2})\sigma _{ij}\right)
\\
& = \frac{1}{N^{2}}\sum_{i=1}^{N}\Vert\gamma _{i}\Vert ^{2}\sigma _{i}^{6}
+\frac{1}{N^{2}}\sum_{i\neq j}\mathrm{tr}%
(\gamma _{i}\gamma _{j}^{\prime })\sigma _{i}^{2}\sigma _{j}^{2}\sigma _{ij}
\\
& \leq \max_{1\leq i\leq N}\Vert \gamma _{i}\Vert^{2}\left(\frac{1}{N}\max_{1\leq i\leq N}\sigma _{i}^{6}+\frac{\Vert \mathcal{E}_{N,T}-\mathrm{dg}(\mathcal{E}%
_{N,T})\Vert }{N}\max_{1\leq i\leq N}\sigma _{i}^{4}\right)\rightarrow 0.
\end{align*}%
Also,
\begin{align*}
\mathrm{tr}\left( \frac{1}{N^{2}}\sum_{i,j=1}^{N}\gamma _{i}\gamma
_{j}^{\prime }\sigma _{ij}^{3}\right) & =\frac{1}{N^{2}}\sum_{i=1}^{N}\Vert
\gamma _{i}\Vert ^{2}\sigma _{i}^{6}+\frac{1}{N^{2}}\sum_{i\neq j}\mathrm{tr}%
(\gamma _{i}\gamma _{j}^{\prime })\sigma _{ij}^{3} \\
& \leq \max_{1\leq i\leq N}\Vert \gamma _{i}\Vert^{2}\left(\frac{1}{N}\max_{1\leq i\leq N}\sigma _{i}^{6}+\frac{1}{N^{2}}\sum_{i,j=1}^{N}\sigma
_{ij}^{2}\max_{1\leq i\leq N}\sigma _{i}^{2}\right)\rightarrow 0.
\end{align*}%
Here we used the statement $N^{-2}\sum_{i,j=1}^{N}\sigma
_{ij}^{2}\rightarrow 0$, which is proved in equation (\ref{eq: prove of
frobenius norm}). This ends the proof of Theorem \ref{th: covariance for
independence}. $\blacksquare $

\subsection{Proofs for Conditional Heteroscedasticity case}

\paragraph{Proof of Theorem \protect\ref{th: CLT for HC}.}

In order to apply Lemma \ref{lem: everything stacked} we check conditions
(i)-(iv) of Lemma \ref{lem: in between lemma} and conditions (a)-(d) of
Lemma \ref{lem: everything stacked} for
\begin{equation*}
W_{st}=\frac{1}{T}w_{st}\frac{e_{t}^{\prime }e_{s}}{\sqrt{N}}
\end{equation*}
and%
\begin{equation*}
V_{s}=\frac{1}{\sqrt{T}}\frac{\gamma ^{\prime }e_{s}}{\sqrt{N}}\otimes v_{s}.
\end{equation*}

\noindent (i) Due to serial independence of $e_{it}$ conditionally on $%
\mathcal{F}$, we have
\begin{equation*}
\Sigma _{W,T}=\frac{1}{T^{2}}\sum_{s=1}^{T}\sum_{t<s}\mathbb{E}\left[
w_{st}w_{st}^{\prime }\mathbb{E}\left( \left( \frac{e_{t}^{\prime }e_{s}}{%
\sqrt{N}}\right) ^{2}|\mathcal{F}\right) \right] .
\end{equation*}%
Notice that $\left( e_{t}^{\prime }e_{s}\right) ^{2}=\mathrm{tr}\left(
\left( e_{s}^{\prime }e_{t}\right) \left( e_{t}^{\prime }e_{s}\right)
\right) =\mathrm{tr}\left( \left( e_{t}e_{t}^{\prime }\right) \left(
e_{s}e_{s}^{\prime }\right) \right) ,$ and hence, given the conditional
independence assumption,%
\begin{equation*}
\mathbb{E}\left[ \left( \frac{e_{t}^{\prime }e_{s}}{\sqrt{N}}\right) ^{2}|%
\mathcal{F}\right] =\frac{1}{N}\mathrm{tr}\big( \mathbb{E}%
(e_{t}e_{t}^{\prime }|\mathcal{F})\mathbb{E}(e_{s}e_{s}^{\prime }|\mathcal{F}%
)\big) .
\end{equation*}%
Recall that $e_{t}=\pi f_{t}+\eta _{t}$. We will use the notation $\Omega
_{\eta }=\mathbb{E}\left( \eta _{t}\eta _{t}^{\prime }\right) =\mathrm{dg}%
\{\omega _{i}^{2}\}_{i=1}^{N}$. Then,%
\begin{align*}
\mathbb{E}\left[ \left( \frac{e_{t}^{\prime }e_{s}}{\sqrt{N}}\right) ^{2}|%
\mathcal{F}\right] & =\frac{1}{N}\mathrm{tr}\big( \left( \pi \mathbb{E}%
(f_{t}f_{t}^{\prime }|\mathcal{F})\pi ^{\prime }+\Omega _{\eta }\right)
\left( \pi \mathbb{E}(f_{s}f_{s}^{\prime }|\mathcal{F})\pi ^{\prime }+\Omega
_{\eta }\right) \big) \\
& =\frac{1}{N}\sum_{i=1}^{N}\omega _{i}^{4}+\Delta _{N,T},
\end{align*}%
where
\begin{equation*}
\Delta _{N,T}\leq \frac{C}{N}\mathbb{E}\left[ (\Vert f_{t}\Vert
^{2}+1)(\Vert f_{s}\Vert ^{2}+1)|\mathcal{F}\right] .
\end{equation*}%
Indeed, $\Delta _{N,T}$ has three terms each of which is easy to bound. For
example,%
\begin{align*}
\frac{1}{N}\mathrm{tr}\big( \Omega _{\eta }\pi \mathbb{E}%
(f_{s}f_{s}^{\prime }|\mathcal{F})\pi ^{\prime }\big) & \leq \frac{1}{N}%
\max_{1\leq i\leq N}\omega _{i}^{2}\cdot \mathrm{tr}\big( \mathbb{E}%
(f_{s}f_{s}^{\prime }|\mathcal{F})\pi ^{\prime }\pi \big) \\
& \leq \frac{1}{N}\max_{1\leq i\leq N}\omega _{i}^{2}\cdot \max \mathrm{ev}%
(\pi ^{\prime }\pi )\cdot \mathbb{E}\left[ \Vert f_{s}\Vert ^{2}|\mathcal{F}%
\right] .
\end{align*}%
Since we assumed that $\max_{1\leq i\leq N}\omega _{i}^{2}<C$ and from
Assumption \ref{ass: heterosk}(ii), it follows that
\begin{equation*}
\left\Vert \frac{1}{T^{2}}\sum_{s=1}^{T}\sum_{t<s}E\left[ w_{st}w_{st}^{%
\prime }\Delta _{N,T}\right] \right\Vert \leq \frac{C}{NT^{2}}%
\sum_{s=1}^{T}\sum_{t<s}E\left[ \Vert w_{st}\Vert ^{2}(\Vert f_{t}\Vert
^{2}+1)(\Vert f_{s}\Vert ^{2}+1)\right] \leq \frac{C}{N}\rightarrow 0,
\end{equation*}%
where the last inequality is due to Assumption \ref{ass: heterosk}(i). So,
we obtain that
\begin{equation*}
\Sigma _{W,T}(T,N)=\lim \frac{1}{T^{2}}\sum_{s=1}^{T}\sum_{t<s}\mathbb{E}%
\left[ w_{st}w_{st}^{\prime }\right] \frac{1}{N}\sum_{i=1}^{N}\omega
_{i}^{4}=\omega ^{4}\Omega _{w}=\Sigma _{W}.
\end{equation*}

\noindent (ii) Notice that
\begin{equation*}
\frac{e_{t}^{\prime }e_{s}}{\sqrt{N}}=\frac{f_{t}^{\prime }\pi ^{\prime }\pi
f_{s}}{\sqrt{N}}+\frac{f_{t}^{\prime }\pi ^{\prime }\eta _{s}}{\sqrt{N}}+%
\frac{f_{s}^{\prime }\pi ^{\prime }\eta _{t}}{\sqrt{N}}+\frac{\eta
_{t}^{\prime }\eta _{s}}{\sqrt{N}}.
\end{equation*}%
Using the Marcinkiewicz--Zygmund inequality for a second power applied twice
we notice that in order to bound $\mathbb{E}\big[ \big( e_{t}^{\prime
}e_{s}/\sqrt{N}\big) ^{4}|\mathcal{F}\big] $ from above it is enough to
bound the fourth moment of each summand. Using serial and cross-sectional
conditional independence of $\eta $'s as well as their conditional
independence from $f$'s, we obtain%
\begin{equation*}
\mathbb{E}\left[ \left( \frac{1}{\sqrt{N}}\sum_{i=1}^{N}\eta _{it}\eta
_{is}\right) ^{4}\right] =\frac{1}{N^{2}}\sum_{i=1}^{N}\mathbb{E}\left[
(\eta _{it}\eta _{is})^{4}\right] +C\frac{1}{N^{2}}\sum_{i_{1}\neq i_{2}}%
\mathbb{E}\big[\eta _{i_{1}t}^{2}\eta _{i_{1}s}^{2}\eta _{i_{2}t}^{2}\eta
_{i_{2}s}^{2}\big]\leq C,
\end{equation*}%
\begin{equation*}
\mathbb{E}\left[ \left\Vert \frac{1}{\sqrt{N}}\sum_{i=1}^{N}\pi _{i}\eta
_{is}\right\Vert ^{4}\right] \leq \frac{1}{N^{2}}\sum_{i=1}^{N}\mathbb{E}%
\left[ \left\Vert \pi _{i}\eta _{is}\right\Vert ^{4}\right] +C\frac{1}{N^{2}}%
\sum_{i_{1}\neq i_{2}}\Vert \pi _{i_{1}}\Vert ^{2}\Vert \pi _{i_{2}}\Vert
^{2}\mathbb{E}\left[ \eta _{i_{1}s}^{2}\eta _{i_{2}s}^{2}\right] \leq \frac{C%
}{N^{2}},
\end{equation*}%
where we use Assumption \ref{ass: heterosk}(ii,iii), and that $\sum_{i}\Vert
\pi _{i}\Vert ^{4}\leq \big(\sum_{i}\Vert \pi _{i}\Vert ^{2}\big)^{2}\leq C$. Hence,
\begin{equation*}
\mathbb{E}\left[ \left( \frac{e_{t}^{\prime }e_{s}}{\sqrt{N}}\right) ^{4}|%
\mathcal{F}\right] \leq \frac{C}{N^{2}}\mathbb{E}\left[ \Vert f_{t}\Vert
^{4}\Vert f_{s}\Vert ^{4}|\mathcal{F}\right] +\frac{C}{N^{2}}\left( \mathbb{E%
}\left[ \Vert f_{t}\Vert ^{4}|\mathcal{F}\right] +E\left[ \Vert f_{s}\Vert
^{4}|\mathcal{F}\right] \right) +C.
\end{equation*}%
Finally, due to Assumption \ref{ass: heterosk}(i),
\begin{equation*}
T^{4}\mathbb{E}\left[ \Vert W_{st}\Vert ^{4}\right] \leq \mathbb{E}\left[
\left\Vert w_{st}\frac{e_{t}^{\prime }e_{s}}{\sqrt{N}}\right\Vert ^{4}\right]
\leq C\mathbb{E}\left[ \Vert w_{st}\Vert ^{4}(\Vert f_{t}\Vert ^{4}+1)(\Vert
f_{s}\Vert ^{4}+1)\right] <C.
\end{equation*}%
Thus, condition (ii) of Lemma \ref{lem: in between lemma} holds.

\noindent (iii) Let us define a $\sigma$-algebra $\mathcal{A}=\mathcal{F}\cup
\{f_{t},t=1,...,T\}$. Let us now denote
\begin{align*}
\vartheta _{st}& =\mathbb{E}\left[ \left( \frac{e_{s}^{\prime }e_{t}}{\sqrt{N%
}}\right) ^{2}|\mathcal{A}\right] =\mathbb{E}\left[ \frac{\big( (\pi
f_{s}+\eta _{s})^{\prime }(\pi f_{t}+\eta _{t})\big) ^{2}}{N}|\mathcal{A}%
\right] \\
& =\frac{1}{N}\left( \left( f_{s}^{\prime }\pi ^{\prime }\pi f_{t}\right)
^{2}+f_{s}^{\prime }\pi \Omega _{\eta }\pi ^{\prime }f_{s}+f_{t}^{\prime
}\pi \Omega _{\eta }\pi ^{\prime }f_{t}+\sum_{i=1}^{N}\omega _{i}^{4}\right)
.
\end{align*}%
We have:
\begin{align*}
\sum_{s=1}^{T}\sum_{t<s}W_{st}W_{st}^{\prime }-\Sigma _{W,T}& =\frac{1}{T^{2}%
}\sum_{s=1}^{T}\sum_{t<s}w_{st}w_{st}^{\prime }\left[ \left( \frac{%
e_{s}^{\prime }e_{t}}{\sqrt{N}}\right) ^{2}-\vartheta _{st}\right] \\
& \quad +\frac{1}{T^{2}}\sum_{s=1}^{T}\sum_{t<s}\big( w_{st}w_{st}^{\prime
}\vartheta _{st}-\mathbb{E}\left[ w_{st}w_{st}^{\prime }\vartheta _{st}%
\right] \big) \\
& =A_{1}+A_{2},
\end{align*}%
so, it is enough to prove convergence of each term separately. Now, $\mathbb{%
E}\left[ \mathrm{tr}(A_{1}A_{1}^{\prime })\right] $ is equal to%
\begin{equation*}
\frac{1}{T^{4}}\sum_{s_{1},s_{2}=1}^{T}\sum_{t_{1},t_{2}}\mathbb{E}\left[
\mathrm{tr}(w_{s_{1}t_{1}}w_{s_{1}t_{1}}^{\prime
}w_{s_{2}t_{2}}w_{s_{2}t_{2}}^{\prime })\left( \left( \frac{%
e_{s_{1}}^{\prime }e_{t_{1}}}{\sqrt{N}}\right) ^{2}-\vartheta
_{s_{1}t_{1}}\right) \left( \left( \frac{e_{s_{2}}^{\prime }e_{t_{2}}}{\sqrt{%
N}}\right) ^{2}-\vartheta _{s_{2}t_{2}}\right) \right] .
\end{equation*}%
Notice that in order for a summand from the last sum to be non-zero we need
that some indexes in the set $\{s_{1},s_{2},t_{1},t_{2}\}$ coincide, and we
obtain at most $CT^{3}$ non-zero summands. Each non-zero summand is bounded
above by a constant due to the moment assumptions formulated in Assumption %
\ref{ass: heterosk}(i,iii). Thus, $\mathbb{E}\left[ \mathrm{tr}%
(A_{1}A_{1}^{\prime })\right] \rightarrow 0$.

Notice that due to Assumption \ref{ass: heterosk}, and similar to the
argument above,
\begin{equation}
\left\vert \vartheta _{st}-\frac{1}{N}\sum_{i=1}^{N}\omega
_{i}^{4}\right\vert \leq \frac{C}{N}(\Vert f_{s}\Vert +\Vert f_{t}\Vert
+1)^{4}.  \label{eq: bound on vareta}
\end{equation}%
Thus,
\begin{align*}
A_{2}& =\left( \frac{1}{N}\sum_{i=1}^{N}\omega _{i}^{4}\right) \frac{1}{T^{2}%
}\sum_{s=1}^{T}\sum_{t<s}\big( w_{st}w_{st}^{\prime }-\mathbb{E}\left(
w_{st}w_{st}^{\prime }\right) \big) \\
& \quad +\,\frac{1}{T^{2}}\sum_{s=1}^{T}\sum_{t<s}\left(
w_{st}w_{st}^{\prime }\big(\vartheta _{st}-\frac{1}{N}\sum_{i=1}^{N}\omega
_{i}^{4}\big)-\mathbb{E}\left[ w_{st}w_{st}^{\prime }\big(\vartheta _{st}-\frac{1}{N}%
\sum_{i=1}^{N}\omega _{i}^{4}\big)\right] \right) ,
\end{align*}%
where the first sum converges to zero due to Assumption \ref{ass: common
vars}(iii), while expectation of the second moment of the second term is
bounded by
\begin{equation*}
\frac{1}{T^{4}}\sum_{s_{1},s_{2}}\sum_{t_{1},t_{2}}\frac{C}{N^{2}}\mathbb{E}%
\big[ (\Vert f_{s_{1}}\Vert +\Vert f_{t_{1}}\Vert +1)^{4}(\Vert
f_{s_{2}}\Vert +\Vert f_{t_{2}}\Vert +1)^{4}\Vert w_{s_{1}t_{1}}\Vert
^{2}\Vert w_{s_{2}t_{2}}\Vert ^{2}\big] \leq \frac{C}{N^{2}},
\end{equation*}%
due to inequality (\ref{eq: bound on vareta}) and Assumption \ref{ass:
heterosk}(i). Thus, condition (iii) of Lemma \ref{lem: in between lemma}
holds.

\noindent Let us check condition (iv):%
\begin{equation*}
T^{4}\mathbb{E}\left( W_{s_{1}t_{2}}^{\prime
}W_{s_{2}t_{1}}W_{s_{2}t_{2}}^{\prime }W_{s_{1}t_{1}}\right) =\frac{1}{N^{2}}%
\mathbb{E}\big[ w_{s_{1}t_{2}}^{\prime
}w_{s_{2}t_{1}}w_{s_{2}t_{2}}^{\prime }w_{s_{1},t_{1}}\mathbb{E}%
(e_{s_{1}}^{\prime }e_{t_{1}}e_{t_{1}}^{\prime }e_{s_{2}}e_{s_{2}}^{\prime
}e_{t_{2}}e_{t_{2}}^{\prime }e_{s_{1}}|\mathcal{F})\big] ,
\end{equation*}%
where we used that the scalar products $e_{t}^{\prime }e_{s}=e_{s}^{\prime
}e_{t}$ are scalars and they can be reshuffled to make two same-index $e_{t}$ 
stand back to back. Let us bound the $N\times N$ matrix $\mathbb{E%
}(e_{t}e_{t}^{\prime }|\mathcal{F})=\pi \mathbb{E}(f_{t}f_{t}^{\prime }|%
\mathcal{F})\pi ^{\prime }+\Omega _{\eta }$:
\begin{align}
\max \mathrm{ev}\big(\mathbb{E}(e_{t}e_{t}^{\prime }|\mathcal{F})\big)& \leq \max
\mathrm{ev}\big(\pi ^{\prime }\mathbb{E}(f_{t}f_{t}^{\prime }|\mathcal{F})\pi
\big)+\max \mathrm{ev}\left( \Omega _{\eta }\right)  \notag \\
& \leq \mathrm{tr}\big(\pi ^{\prime }\mathbb{E}(f_{t}f_{t}^{\prime }|\mathcal{F}%
)\pi \big)+\max_{1\leq i\leq N}\omega _{i}^{2}  \notag \\
& \leq \max \mathrm{ev}(\pi \pi ^{\prime })\mathbb{E}\big(\Vert f_{t}\Vert ^{2}|%
\mathcal{F}\big)+C  \notag \\
& \leq C\mathbb{E}\big(\Vert f_{t}\Vert ^{2}+1|\mathcal{F}\big).  \label{eq: bound 1}
\end{align}%
As a result,
\begin{align*}
\left\vert \mathbb{E}(e_{s_{1}}^{\prime }e_{t_{1}}e_{t_{1}}^{\prime
}e_{s_{2}}e_{s_{2}}^{\prime }e_{t_{2}}e_{t_{2}}^{\prime }e_{s_{1}}|\mathcal{F%
})\right\vert & =\left\vert \mathrm{tr}\big(\mathbb{E}(e_{t_{1}}e_{t_{1}}^{%
\prime }|\mathcal{F})\mathbb{E}(e_{s_{2}}e_{s_{2}}^{\prime }|\mathcal{F})%
\mathbb{E}(e_{t_{2}}e_{t_{2}}^{\prime }|\mathcal{F})\mathbb{E}%
(e_{s_{1}}e_{s_{1}}^{\prime }|\mathcal{F})\big)\right\vert \\
& \leq N\max \mathrm{ev}\left(\prod_{t\in \{s_{1},s_{2},t_{1},t_{2}\}}\mathbb{E}%
(e_{t}e_{t}^{\prime }|\mathcal{F})\right) \\
& \leq N\prod_{t\in \{s_{1},s_{2},t_{1},t_{2}\}}\max \mathrm{ev}\left(
\mathbb{E}(e_{t}e_{t}^{\prime }|\mathcal{F})\right) \\
& \leq NC\prod_{t\in \{s_{1},s_{2},t_{1},t_{2}\}}\mathbb{E}\big( \Vert
f_{t}\Vert ^{2}+1|\mathcal{F}\big) .
\end{align*}%
Also using Assumption \ref{ass: heterosk}(i) we obtain that%
\begin{equation*}
T^{4}\left\vert E\left( W_{s_{1}t_{2}}^{\prime
}W_{s_{2}t_{1}}W_{s_{2}t_{2}}^{\prime }W_{s_{1}t_{1}}\right) \right\vert
\leq \frac{C}{N}\max_{1\leq s,t,t^{\ast }\leq T}E\left[ \Vert w_{st}\Vert
^{4}\Vert f_{t^{\ast }}\Vert ^{8}\right] \rightarrow 0.
\end{equation*}%
Thus, condition (iv) holds as well.

\noindent Now we will check assumptions (a)-(d) of Lemma \ref{lem:
everything stacked}. First, we find the limit of the covariance matrix $%
\Sigma _{V,T}$.
\begin{align*}
\mathbb{E}\left[ \left( \frac{\gamma ^{\prime }e_{s}}{\sqrt{N}}\right)
\left( \frac{\gamma ^{\prime }e_{s}}{\sqrt{N}}\right) ^{\prime }|\mathcal{F}%
\right] & =\left( \frac{\gamma ^{\prime }\pi }{\sqrt{N}}\right) \mathbb{E}%
[f_{s}f_{s}^{\prime }|\mathcal{F}]\left( \frac{\pi ^{\prime }\gamma }{\sqrt{N%
}}\right) +\frac{1}{N}\sum_{i=1}^{N}\omega _{i}^{2}\gamma _{i}\gamma
_{i}^{\prime } \\
& \rightarrow \Gamma _{\pi \gamma }^{\prime }\mathbb{E}\left(
f_{s}f_{s}^{\prime }|\mathcal{F}\right) \Gamma _{\pi \gamma }+\Gamma
_{\omega }.
\end{align*}%
Here we used Assumptions \ref{ass: heterosk}(ii,iii). Therefore,
\begin{eqnarray*}
\Sigma _{V,T} &=&\mathrm{var}\left( \sum_{s=1}^{T}V_{s}\right) =\mathbb{E}%
\left[ \frac{1}{T}\sum_{s=1}^{T}\mathbb{E}\left( \left( \frac{\gamma
^{\prime }e_{s}}{\sqrt{N}}\right) \left( \frac{\gamma ^{\prime }e_{s}}{\sqrt{%
N}}\right) ^{\prime }|\mathcal{F}\right) \otimes \left( v_{s}v_{s}^{\prime
}\right) \right] \\
&=&\frac{1}{T}\sum_{s=1}^{T}\mathbb{E}\left[ \left( \Gamma _{\pi \gamma
}^{\prime }f_{s}f_{s}^{\prime }\Gamma _{\pi \gamma }+\Gamma _{\omega
}\right) \otimes \left( v_{s}v_{s}^{\prime }\right) \right] \\
&\rightarrow &(\Gamma _{\pi \gamma }^{\prime }\otimes I_{k_{v}})\Sigma
_{fv}(\Gamma _{\pi \gamma }\otimes I_{k_{v}})+\Gamma _{\omega }\otimes
\Omega _{v}.
\end{eqnarray*}%
The limit matrix is positive definite since both $\Gamma _{\omega }$ and $%
\Omega _{v}$ are positive-definite due to Assumptions \ref{ass: common vars}%
(i) and \ref{ass: heterosk}(iii).

\noindent Now note that due to Assumption \ref{ass: heterosk}(ii),
\begin{equation*}
\mathbb{E}\left[ \left\Vert \frac{\gamma ^{\prime }e_{t}}{\sqrt{N}}%
\right\Vert ^{4}|\mathcal{F}\right] =\frac{1}{N^{2}}\mathbb{E}\left( \Vert
\gamma ^{\prime }\pi f_{t}+\gamma ^{\prime }\eta _{t}\Vert ^{4}|\mathcal{F}%
\right) \leq C\mathbb{E}\left( \Vert f_{t}\Vert ^{4}|\mathcal{F}\right) +%
\frac{C}{N^{2}}\mathbb{E}\left( \Vert \gamma ^{\prime }\eta _{t}\Vert
^{4}\right) ,
\end{equation*}%
\begin{equation*}
\mathbb{E}\left( \Vert \gamma ^{\prime }\eta _{t}\Vert ^{4}\right) =\mathbb{E%
}\left[ \left\Vert \sum_{i=1}^{N}\gamma _{i}^{\prime }\eta _{it}\right\Vert
^{4}\right] \leq \sum_{i=1}^{N}\Vert \gamma _{i}\Vert ^{4}\mathbb{E}(\eta
_{it}^{4})+C\sum_{i_{1},i_{2}=1}^{N}\Vert \gamma _{i_{1}}\Vert ^{2}\Vert
\gamma _{i_{2}}\Vert ^{2}\omega _{i_{1}}^{2}\omega _{i_{2}}^{2}.
\end{equation*}%
Due to Assumptions \ref{ass: loadings} and \ref{ass: heterosk}(iii) we have
that $\mathbb{E}\big( \Vert \gamma ^{\prime }\eta _{t}\Vert ^{4}\big)
\leq CN^{2},$ and thus
\begin{equation*}
\mathbb{E}\left[ \left\Vert \frac{\gamma ^{\prime }e_{t}}{\sqrt{N}}%
\right\Vert ^{4}|\mathcal{F}\right] \leq C\mathbb{E}\left( \Vert f_{t}\Vert
^{4}+1|\mathcal{F}\right) .
\end{equation*}%
Collecting the pieces,%
\begin{equation*}
T\mathbb{E}\big(\Vert V_{s}\Vert ^{4}\big)\leq CT\mathbb{E}\left[ \frac{1}{T^{2}}%
\mathbb{E}\left[ \left\Vert \frac{\gamma ^{\prime }e_{s}}{\sqrt{N}}%
\right\Vert ^{4}|\mathcal{F}\right] \otimes \Vert v_{s}\Vert ^{4}\right]
\leq \frac{C}{T}\mathbb{E}\left[ \left( \Vert f_{s}\Vert ^{4}+1\right)
\Vert v_{s}\Vert ^{4}\right] \rightarrow 0.
\end{equation*}%
This gives us the validity of condition (b) of Lemma \ref{lem: everything
stacked}.

\noindent (c) Denote $\Gamma _{\omega ,N}=N^{-1}\sum_{i=1}^{N}\omega
_{i}^{2}\gamma _{i}\gamma _{i}^{\prime }\rightarrow \Gamma _{\omega }.$
Then,
\begin{align*}
\sum_{t=1}^{T}V_{t}V_{t}^{\prime }-\Sigma _{V,T}& =\frac{1}{T}%
\sum_{t=1}^{T}\left( \frac{\gamma ^{\prime }e_{t}}{\sqrt{N}}\frac{%
e_{t}^{\prime }\gamma }{\sqrt{N}}\right) \otimes (v_{t}v_{t}^{\prime }) \\
& \quad -\frac{1}{T}\sum_{t=1}^{T}\mathbb{E}\left[ \left( \frac{\gamma
^{\prime }\pi }{\sqrt{N}}f_{t}f_{t}^{\prime }\frac{\pi ^{\prime }\gamma }{%
\sqrt{N}}+\Gamma _{\omega ,N}\right) \otimes \left( v_{t}v_{t}^{\prime
}\right) \right] \\
& =\frac{1}{T}\sum_{t=1}^{T}\left( \frac{\gamma ^{\prime }e_{t}}{\sqrt{N}}
\frac{e_{t}^{\prime }\gamma }{\sqrt{N}}-\frac{\gamma ^{\prime }\pi }{\sqrt{N}%
}f_{t}f_{t}^{\prime }\frac{\pi ^{\prime }\gamma }{\sqrt{N}}-\Gamma _{\omega
,N}\right) \otimes (v_{t}v_{t}^{\prime }) \\
& \quad +\frac{1}{T}\sum_{t=1}^{T}\left[ \left( \frac{\gamma ^{\prime }\pi }{%
\sqrt{N}}f_{t}f_{t}^{\prime }\frac{\pi ^{\prime }\gamma }{\sqrt{N}}+\Gamma
_{\omega ,N}-\mathbb{E}\left[ \frac{\gamma ^{\prime }\pi }{\sqrt{N}}%
f_{t}f_{t}^{\prime }\frac{\pi ^{\prime }\gamma }{\sqrt{N}}+\Gamma _{\omega
,N}\right] \right) \otimes \left( v_{t}v_{t}^{\prime }\right) \right] \\
& =A_{1}+A_{2}.
\end{align*}%
Notice that given the conditional independence of $\eta _{it}$'s, the two
terms in the last expression, $A_{1}$ and $A_{2}$ are uncorrelated, so in
order to check condition (c) of Lemma \ref{lem: everything stacked} we can
prove convergence to zero of $\mathbb{E}\big[\Vert A_{1}\Vert ^{2}\big]$ and $\mathbb{E%
}\big[\Vert A_{2}\Vert ^{2}\big]$ separately. First,
\begin{align*}
\mathbb{E}\left[ \Vert A_{1}\Vert ^{2}\right] & =\mathbb{E}\left[ \left\Vert
\frac{1}{T}\sum_{t=1}^{T}\left( \frac{\gamma ^{\prime }\pi }{\sqrt{N}}f_{t}%
\frac{\eta _{t}^{\prime }\gamma }{\sqrt{N}}+\frac{\gamma ^{\prime }\eta _{t}%
}{\sqrt{N}}f_{t}^{\prime }\frac{\pi ^{\prime }\gamma }{\sqrt{N}}+\left(
\frac{\gamma ^{\prime }\eta _{t}}{\sqrt{N}}\frac{\eta _{t}^{\prime }\gamma }{%
\sqrt{N}}-\Gamma _{\omega ,N}\right) \right) \otimes \left(
v_{t}v_{t}^{\prime }\right) \right\Vert ^{2}\right] \\
& \leq \frac{1}{T^{2}}\sum_{t=1}^{T}\mathbb{E}\left[ \left\Vert \frac{\gamma
^{\prime }\pi }{\sqrt{N}}f_{t}\frac{\eta _{t}^{\prime }\gamma }{\sqrt{N}}+%
\frac{\gamma ^{\prime }\eta _{t}}{\sqrt{N}}f_{t}^{\prime }\frac{\pi ^{\prime
}\gamma }{\sqrt{N}}+\left( \frac{\gamma ^{\prime }\eta _{t}}{\sqrt{N}}\frac{%
\eta _{t}^{\prime }\gamma }{\sqrt{N}}-\Gamma _{\omega ,N}\right) \right\Vert
^{2}\Vert v_{t}\Vert ^{4}\right] \\
& \leq \frac{1}{T}C\mathbb{E}\left[ (\Vert f_{t}\Vert ^{2}+1)\Vert
v_{t}\Vert ^{4}\right] \rightarrow 0.
\end{align*}%
The former inequality is due to $\eta _{t}$'s being conditionally serially
uncorrelated, and thus the summation over $t$ can be taken outside the
expectation of the square; the latter inequality uses bounds on the moments
of $\eta _{t}^{\prime }\gamma /\sqrt{N}$ we derived before. Second, the
convergence of term $A_{2}$ is due to Assumptions \ref{ass: heterosk}(iv)
and \ref{ass: common vars}(iv). Putting all terms together, we obtain that
condition (c) is satisfied.

\noindent Finally, we check the condition (d):
\begin{equation*}
T^{3}\left\Vert \mathbb{E}\left( W_{s_{1}t}V_{s_{1}}^{\prime
}V_{s_{2}}W_{s_{2}t}^{\prime }\right) \right\Vert =\left\Vert \mathbb{E}%
\left[ w_{s_{1}t}v_{s_{1}}^{\prime }v_{s_{2}}w_{s_{2}t}^{\prime }\mathbb{E}%
\left( \frac{e_{s_{1}}^{\prime }\gamma }{\sqrt{N}}\frac{\gamma ^{\prime
}e_{s_{2}}}{\sqrt{N}}\frac{e_{s_{1}}^{\prime }e_{t}}{\sqrt{N}}\frac{%
e_{s_{2}}^{\prime }e_{t}}{\sqrt{N}}|\mathcal{F}\right) \right] \right\Vert .
\end{equation*}%
Using that scalars could be reshuffled to make two same-index $e_{t}$ 
stand back to back and employing conditional independence we obtain:%
\begin{align*}
\left\vert \mathbb{E}\left( \frac{e_{s_{1}}^{\prime }\gamma }{\sqrt{N}}\frac{%
\gamma ^{\prime }e_{s_{2}}}{\sqrt{N}}\frac{e_{s_{1}}^{\prime }e_{t}}{\sqrt{N}%
}\frac{e_{s_{2}}^{\prime }e_{t}}{\sqrt{N}}|\mathcal{F}\right) \right\vert & =%
\frac{1}{N^{2}}\left\vert \mathrm{tr}\big(\gamma \gamma ^{\prime }\mathbb{E}%
(e_{s_{2}}e_{s_{2}}^{\prime }|\mathcal{F})\mathbb{E}(e_{t}e_{t}^{\prime }|%
\mathcal{F})\mathbb{E}(e_{s_{1}}e_{s_{1}}^{\prime }|\mathcal{F})\big)\right\vert
\\
& \leq \frac{1}{N^{2}}\mathrm{tr}(\gamma \gamma ^{\prime })\prod_{s\in
\{s_{1},s_{2},t\}}\max \mathrm{ev}\big(\mathbb{E}(e_{s}e_{s}^{\prime }|\mathcal{F%
})\big) \\
& \leq \frac{C}{N}\mathbb{E}\left[ \prod_{s\in \{s_{1},s_{2},t\}}\big(\Vert
f_{s}\Vert ^{2}+1\big)|\mathcal{F}\right] .
\end{align*}%
We use Assumption \ref{ass: loadings} that $N^{-1}\mathrm{tr}(\gamma \gamma
^{\prime })<C$ and the bound (\ref{eq: bound 1}) we derived before. In the
last equality, we also exploit that $f_{t}$'s are conditionally independent
of each other. Thus, Assumption \ref{ass: heterosk} (i) implies that
\begin{equation*}
T^{3}\max_{1\leq t<\min \{s_{1},s_{2}\}\leq T}\left\Vert E\left(
W_{s_{1}t}V_{s_{1}}^{\prime }V_{s_{2}}W_{s_{2}t}^{\prime }\right)
\right\Vert \leq \frac{C}{N}\rightarrow 0.
\end{equation*}%
Thus, condition (d) of Lemma \ref{lem: everything stacked} is satisfied.
This concludes the proof of Theorem \ref{th: CLT for HC}. $\blacksquare $

\paragraph{Proof of Theorem \protect\ref{th: covariance for HC}.}

We will prove three statements:

\begin{itemize}
\item[(i)] $N^{-1}\sum_{i=1}^{N}\xi _{V,i}\xi _{V,i}^{\prime }\rightarrow
\Sigma _{V}$;

\item[(ii)] $N^{-1}\sum_{i=1}^{N}\xi _{W,i}\xi _{W,i}^{\prime }\rightarrow
\Sigma _{W}$;

\item[(iii)] $N^{-1}\sum_{i=1}^{N}\xi _{V,i}\xi _{W,i}^{\prime }\rightarrow
0 $.
\end{itemize}

\noindent (i) Given assumption $\Gamma _{\pi \gamma }=0$, we have $\Sigma
_{V}=\Gamma _{\omega }\otimes \Omega _{v}.$ Then,
\begin{equation}
\frac{1}{N}\sum_{i=1}^{N}\xi _{V,i}\xi _{V,i}^{\prime }=\frac{1}{NT}%
\sum_{t=1}^{T}\sum_{s=1}^{T}\left( \sum_{i=1}^{N}\gamma _{i}\gamma
_{i}^{\prime }(\pi _{i}^{\prime }f_{s}+\eta _{is})(\pi _{i}^{\prime
}f_{t}+\eta _{it})\right) \otimes (v_{s}v_{t}^{\prime }).
\label{eq: another long}
\end{equation}%
After we open up the brackets there will be three different types of terms.
We will show that
\begin{equation}
\frac{1}{NT}\sum_{t=1}^{T}\sum_{s=1}^{T}\sum_{i=1}^{N}\left( \gamma
_{i}\gamma _{i}^{\prime }\eta _{is}\eta _{it}\right) \otimes \left(
v_{s}v_{t}^{\prime }\right) \overset{p}{\rightarrow }\Sigma _{V},
\label{eq: main term in sigmaV}
\end{equation}%
while terms that involve $\pi _{i}^{\prime }w_{s}\pi _{i}^{\prime }w_{t}$ or
$\eta _{it}\pi _{i}^{\prime }f_{s}$ converge to zero in probability. Indeed,%
\begin{eqnarray*}
\frac{1}{NT}\sum_{t=1}^{T}\sum_{s=1}^{T}\sum_{i=1}^{N}\left( \gamma
_{i}\gamma _{i}^{\prime }\eta _{is}\eta _{it}\right) \otimes
(v_{s}v_{t}^{\prime })-\Sigma _{V,T}\!\!\! &=&\!\!\!\frac{1}{NT}%
\sum_{t=1}^{T}\sum_{s\neq t}^{T}\sum_{i=1}^{N}\left( \gamma _{i}\gamma
_{i}^{\prime }\eta _{is}\eta _{it}\right) \otimes (v_{s}v_{t}^{\prime }) \\
&&\!\!\!+\,\frac{1}{NT}\sum_{t=1}^{T}\sum_{i=1}^{N}\left( \gamma _{i}\gamma
_{i}^{\prime }\right) \otimes \big(\eta _{it}^{2}v_{t}v_{t}^{\prime }-\omega
_{i}^{2}\mathbb{E}\left( v_{t}v_{t}^{\prime }\right) \!\big).
\end{eqnarray*}%
We check that the first sum in the last expression is negligible:
\begin{eqnarray*}
\mathbb{E}\left[ \mathrm{tr}\left( \left( \frac{1}{NT}\sum_{t=1}^{T}\sum_{s%
\neq t}^{T}\sum_{i=1}^{N}\left( \gamma _{i}\gamma _{i}^{\prime }\eta
_{is}\eta _{it}\right) \otimes (v_{s}v_{t}^{\prime })\right)^{\!2}\right) %
\right] \!\!\! &\leq &\!\!\!\frac{1}{N^{2}T^{2}}\sum_{t=1}^{T}\sum_{s\neq
t}^{T}\sum_{i=1}^{N}\Vert \gamma _{i}\Vert ^{4}\omega _{i}^{4}\mathbb{E}%
\big[\Vert v_{t}\Vert ^{2}\Vert v_{s}\Vert ^{2}\big] \\
\!\!\! &\leq &\!\!\!\frac{C}{N^{2}}\sum_{i=1}^{N}\Vert \gamma _{i}\Vert
^{4}\omega _{i}^{4}\rightarrow 0.
\end{eqnarray*}%
Here we use the conditional cross-sectional and temporal independence of $%
\eta _{it}$, that is, for $s\neq t$ we have $\mathbb{E}(\eta _{it}\eta
_{is}\eta _{jt^{\ast }}\eta _{js^{\ast }}|\mathcal{F})=\omega _{i}^{4}$ if $%
i=j$ and $\{t,s\}=\{t^{\ast },s^{\ast }\}$, and zero otherwise. We also use
Assumptions \ref{ass: loadings} and \ref{ass: heterosk}(iii). As for the
second sum, we notice that all summands in the expression below are
uncorrelated with each other, hence
\begin{align*}
& \mathrm{tr}\left( \mathbb{E}\left[ \left( \frac{1}{NT}\sum_{i=1}^{N}%
\sum_{t=1}^{T}\left( \gamma _{i}\gamma _{i}^{\prime }\eta _{it}^{2}\right)
\otimes (v_{t}v_{t}^{\prime })-\Sigma _{V}\right) ^{2}\right] \right) \\
& =\frac{1}{N^{2}T^{2}}\sum_{i=1}^{N}\sum_{t=1}^{T}\mathrm{tr}\left( \mathbb{%
E}\left[ \left( \left( \gamma _{i}\gamma _{i}^{\prime }\eta _{it}^{2}\right)
\otimes (v_{t}v_{t}^{\prime })-\mathbb{E}\left[ \left( \gamma _{i}\gamma
_{i}^{\prime }\eta _{it}^{2}\right) \otimes (v_{t}v_{t}^{\prime })\right]
\right) ^{2}\right] \right) \\
& \leq \frac{C}{N^{2}T^{2}}\sum_{i=1}^{N}\sum_{t=1}^{T}\Vert \gamma
_{i}\Vert ^{4}\mathbb{E}\left[ \Vert v_{t}\Vert ^{4}\right] \rightarrow 0.
\end{align*}%
Thus, we showed the convergence (\ref{eq: main term in sigmaV}).

Now consider terms in (\ref{eq: another long}) that involve $\pi
_{i}^{\prime }f_{s}\pi _{i}^{\prime }f_{t}$:
\begin{align*}
& \frac{1}{NT}\sum_{t=1}^{T}\sum_{s=1}^{T}\left( \sum_{i=1}^{N}\gamma
_{i}\gamma _{i}^{\prime }\pi _{i}^{\prime }f_{s}\pi _{i}^{\prime
}f_{t}\right) \otimes (v_{s}v_{t}^{\prime }) \\
& \quad=\left( \frac{1}{N}\sum_{i=1}^{N}\left( \gamma _{i}\gamma _{i}^{\prime
}\right) \otimes (\pi _{i}^{\prime }\otimes \pi _{i}^{\prime })\otimes
I_{k_{v}}\right) \left( \frac{1}{T}\sum_{t=1}^{T}\sum_{s=1}^{T}I_{k_{\gamma
}}\otimes vec(f_{s}f_{t}^{\prime })\otimes (v_{s}v_{t}^{\prime })\right) .
\end{align*}%
Using Assumption \ref{ass: loadings} and \ref{ass: heterosk}(ii) we can show
that
\begin{equation*}
\left\Vert \frac{1}{N}\sum_{i=1}^{N}\left( \gamma _{i}\gamma _{i}^{\prime
}\right) \otimes \left( \pi _{i}^{\prime }\otimes \pi _{i}^{\prime }\right)
\right\Vert \leq \frac{1}{N}\sum_{i=1}^{N}\left\Vert \gamma _{i}\right\Vert
^{2}\left\Vert \pi _{i}\right\Vert ^{2}\leq C\frac{1}{N}\sum_{i=1}^{N}\left%
\Vert \pi _{i}\right\Vert ^{2}\rightarrow 0.
\end{equation*}%
Now observe that
\begin{align*}
& \mathbb{E}\left[ \left\Vert \frac{1}{T}\sum_{t=1}^{T}%
\sum_{s=1}^{T}vec(f_{s}f_{t}^{\prime })\otimes (v_{s}v_{t}^{\prime
})\right\Vert _{F}^{2}\right] \\
& \quad=\mathrm{tr}\left( \frac{1}{T^{2}}\mathbb{E}\left[ \sum_{t=1}^{T}%
\sum_{s=1}^{T}\sum_{t^{\ast }=1}^{T}\sum_{s^{\ast }=1}^{T}\big(
vec(f_{s}f_{t}^{\prime })\;vec(f_{s^{\ast }}f_{t^{\ast }}^{\prime })^{\prime
}\big) \otimes (v_{s}v_{t}^{\prime }v_{s^{\ast }}v_{t^{\ast }}^{\prime })%
\right] \right) \\
& \quad\leq C\frac{1}{T^{2}}\mathbb{E}\left[ \sum_{t=1}^{T}\sum_{s=1}^{T}\Vert
f_{t}\Vert ^{2}\Vert f_{s}\Vert ^{2}\Vert v_{s}\Vert ^{2}\Vert v_{t}\Vert
^{2}\right] <C.
\end{align*}%
Here the equality is due to $f_{t}$'s being serially independent and mean
zero conditionally on $\mathcal{F}$ by Assumption \ref{ass: heterosk}(i) and
$v_{t}\in \mathcal{F}$; hence, among the four summation indexes at most two
may be distinct. The last inequality is due to Assumption \ref{ass: heterosk}%
(i). Thus, we showed that%
\begin{equation*}
\frac{1}{NT}\sum_{t=1}^{T}\sum_{s=1}^{T}\left( \sum_{i=1}^{N}\gamma
_{i}\gamma _{i}^{\prime }\pi _{i}^{\prime }f_{s}\pi _{i}^{\prime
}f_{t}\right) \otimes \left( v_{s}v_{t}^{\prime }\right) \overset{p}{%
\rightarrow }0.
\end{equation*}%
And finally, we show that
\begin{equation*}
\frac{1}{NT}\sum_{t=1}^{T}\sum_{s=1}^{T}\left( \sum_{i=1}^{N}\gamma
_{i}\gamma _{i}^{\prime }\pi _{i}^{\prime }f_{s}\eta _{it}\right) \otimes
\left( v_{s}v_{t}^{\prime }\right) \overset{p}{\rightarrow }0.
\end{equation*}%
This holds because $\eta _{it}$'s are mean zero, cross-sectionally
independent and independent from $f_{t}$ conditionally on $\mathcal{F}$.
This implies that the mean of the sum above is zero, and all summands are
uncorrelated with each other. The second moment of the sum is bounded above
by
\begin{equation*}
\frac{C}{N^{2}T^{2}}\sum_{t=1}^{T}\sum_{s=1}^{T}\sum_{i=1}^{N}\Vert \gamma
_{i}\Vert ^{4}\Vert \pi _{i}\Vert ^{2}\omega _{i}^{2}\mathbb{E}\left[ \Vert
f_{t}\Vert ^{2}\Vert v_{t}\Vert ^{2}\Vert v_{s}\Vert ^{2}\right] \rightarrow
0.
\end{equation*}%
Thus, we proved statement (i).

Let us turn to statement (ii):
\begin{align*}
\frac{1}{N}\sum_{i=1}^{N}\xi _{W,i}\xi _{W,i}^{\prime }-\Sigma _{W,T}& =%
\frac{1}{T^{2}N}\sum_{i=1}^{N}\sum_{s=1}^{T}\sum_{t<s}w_{st}w_{st}^{\prime
}\left( e_{it}^{2}e_{is}^{2}-\omega _{i}^{4}\right) \\
& \quad +\frac{1}{T^{2}N}\sum_{i=1}^{N}\sum_{s=1}^{T}\sum_{t<s}\sum_{s^{\ast
}=1}^{T}\sum_{t^{\ast }<s\ast ,\{s,t\}\neq \{s^{\ast },t^{\ast
}\}}w_{st}w_{s^{\ast }t^{\ast }}^{\prime }e_{it}e_{is}e_{it^{\ast
}}e_{is^{\ast }} \\
& \quad +\frac{1}{N}\sum_{i=1}^{N}\omega _{i}^{4}\frac{1}{T^{2}}%
\sum_{s=1}^{T}\sum_{t<s}\big( w_{st}w_{st}^{\prime }-\mathbb{E}%
(w_{st}w_{st}^{\prime })\big) \\
& =A_{1}+A_{2}+A_{3}.
\end{align*}%
As for $A_{1}$, we can notice that all summands with indexes $\{s,t\}\neq
\{s^{\ast },t^{\ast }\}$ are uncorrelated with each other, so the
correlation for summands with different indexes $i$ can come only from the $%
\pi _{i}^{\prime }f_{t}$ part. Thus,
\begin{align*}
\mathbb{E}\big[ \left\Vert A_{1}\right\Vert _{F}^{2}\big] & =\frac{1}{%
T^{4}N^{2}}\sum_{s=1}^{T}\sum_{t<s}\mathbb{E}\left[ \left\Vert
\sum_{i=1}^{N}w_{st}w_{st}^{\prime }\left( e_{it}^{2}e_{is}^{2}-\omega
_{i}^{4}\right) \right\Vert _{F}^{2}\right] \\
& \leq \frac{C}{T^{4}N^{2}}\sum_{s=1}^{T}\sum_{t<s}\sum_{i=1}^{N}\left(
\begin{array}{l}
\mathbb{E}\big[ \Vert w_{st}\Vert ^{4}\big] \max_{1\leq i\leq N,1\leq
t\leq T}\mathbb{E}\big( \eta _{it}^{4}\big) ^{2} \\
\quad +\,\sum_{j=1}^{N}\Vert \pi _{i}\Vert ^{4}\Vert \pi _{j}\Vert ^{4}%
\mathbb{E}\big[ \Vert w_{st}\Vert ^{4}\Vert f_{t}\Vert ^{4}\Vert f_{s}\Vert
^{4}\big]%
\end{array}%
\right) \\
& \rightarrow 0.
\end{align*}%
In the last convergence we used that due to Assumption \ref{ass: heterosk},
\begin{equation}
\frac{C}{N^{2}}\sum_{i=1}^{N}\sum_{j=1}^{N}\Vert \pi _{i}\Vert ^{4}\Vert \pi
_{j}\Vert ^{4}\leq \frac{C}{N^{2}}\max_{1\leq i\leq N}\Vert \pi _{i}\Vert
^{4}\left( \sum_{i=1}^{N}\Vert \pi _{i}\Vert ^{2}\right) ^{2}\rightarrow 0,
\label{eq: bounds on pi^4}
\end{equation}%
and hence the term $A_{1}$ converges to zero.

Term $\mathbb{E}\big[ \Vert A_{2}\Vert _{F}^{2}\big] $ equals the
following expression:
\begin{equation}
\frac{1}{T^{4}N^{2}}\sum_{i,j=1}^{N}\sum_{\substack{ s_{1},s_{1}^{\ast },
\\ s_{2},s_{2}^{\ast }}}^{T}\sum_{\substack{ t_{m}<s_{m},  \\ t_{m}^{\ast
}<s_{m}^{\ast }  \\ \{s_{m},t_{m}\}\neq \{s_{m}^{\ast },t_{m}^{\ast }\}}}%
\mathbb{E}\big[ \mathrm{tr}(w_{s_{1}t_{1}}w_{s_{1}^{\ast }t_{1}^{\ast
}}^{\prime }w_{s_{2}t_{2}}w_{s_{2}^{\ast }t_{2}^{\ast }}^{\prime
})e_{it_{1}}e_{is_{1}}e_{it_{1}^{\ast }}e_{is_{1}^{\ast
}}e_{jt_{2}}e_{js_{2}}e_{jt_{2}^{\ast }}e_{js_{2}^{\ast }}\big] .
\label{eq: looong sum}
\end{equation}%
Notice that if $s_{m}<t_{m}$, $s_{m}^{\ast }<t_{m}^{\ast }$ and $%
\{s_{m},t_{m}\}\neq \{s_{m}^{\ast },t_{m}^{\ast }\}$ for $m=1,2,$ the only
ways when the expectation%
\begin{equation}
\mathbb{E}(e_{it_{1}}e_{is_{1}}e_{it_{1}^{\ast }}e_{is_{1}^{\ast
}}e_{jt_{2}}e_{js_{2}}e_{jt_{2}^{\ast }}e_{js_{2}^{\ast }}|\mathcal{F})\neq 0
\label{eq: indexes}
\end{equation}%
can be non zero is when we place at least four restrictions on the time
indexes. Indeed, if $\{s_{1},s_{1}^{\ast },t_{1},t_{1}^{\ast }\}$ are all
distinct, then to get a non zero expectation we need indexes to coincide as
sets: $\{s_{1},s_{1}^{\ast },t_{1},t_{1}^{\ast }\}=\{s_{2},s_{2}^{\ast
},t_{2},t_{2}^{\ast }\}$. If the set $\{s_{1},s_{1}^{\ast
},t_{1},t_{1}^{\ast }\}$ contains three distinct indexes, for example, $%
s_{1}=s_{1}^{\ast }$ (this is one restriction), then the set $%
\{s_{2},s_{2}^{\ast },t_{2},t_{2}^{\ast }\}$ should contain $%
(t_{1},t_{1}^{\ast })$ (these are two restrictions), and the remaining
indexes should be either equal to each other (one restriction) or equal to
the ones previously mentioned (two restrictions). Thus, instead of
eight-dimensional summation over time indexes in equation (\ref{eq: looong
sum}) we have a four-dimensional summation.

Let us consider those terms in (\ref{eq: looong sum}) when the summation
index $j$ is equal to $i$. Notice that since each $t$ index is strictly
smaller than the corresponding $s$ index, then any distinct time index can
appear in the set $\{s_{1},s_{1}^{\ast },t_{1},t_{1}^{\ast
},s_{2},s_{2}^{\ast },t_{2},t_{2}^{\ast }\}$ at most four times, thus any
individual error term $e_{it}$ may appear in at most power four. Thus, all
non-zero terms are bounded above by $\max_{1\leq i\leq N,1\leq s,t,t^{\ast
}\leq T}\mathbb{E}\big[ \Vert w_{st}\Vert ^{4}\big( \mathbb{E}\left( \eta
_{it}^{4}\right) +C\Vert f_{t^{\ast }}\Vert ^{4}\big) ^{2}\big] <C$ due
to Assumption \ref{ass: heterosk}(i,iii). There are at most $CT^{4}N$ of
such terms while the normalization is $N^{-2}T^{-4}$, hence that sum
converges to zero.

Now consider those terms in (\ref{eq: looong sum}) when $i\neq j$. Since $%
e_{it}=\pi _{i}^{\prime }f_{t}+\eta _{it},$ with $\eta _{it}$'s independent
of each other both cross-sectionally and temporally, $i\neq j$ and $%
\{s_{m},t_{m}\}\neq \{s_{m}^{\ast },t_{m}^{\ast }\}$, we have that all terms
including $\eta _{it}$ are zero, and only a non-trivial part of the term in (%
\ref{eq: indexes}) is the one including $\pi _{i}^{\prime }f_{t}$ in place
of $e_{it}$. So, every non-zeros term in the sum (\ref{eq: looong sum}) is
bounded by $\Vert \pi _{i}\Vert ^{4}\Vert \pi _{j}\Vert ^{4}\mathbb{E}\big[
\Vert w_{st}\Vert ^{4}\Vert f_{t}\Vert ^{8}\big] $. So, the sum in (\ref%
{eq: looong sum}) over $j\neq i$ is bounded above in the same manner as
stated in equation (\ref{eq: bounds on pi^4}). Thus, we showed that $A_{2}%
\overset{p}{\rightarrow }0.$ The convergence $A_{3}\overset{p}{\rightarrow }%
0 $ comes from Assumption \ref{ass: common vars}(iii). This finishes the
proof of (ii).

Finally, let us prove statement (iii):
\begin{equation*}
\frac{1}{NT^{3/2}}\sum_{i=1}^{N}\sum_{s=1}^{T}\sum_{t<s}\sum_{s^{\ast
}=1}^{T}(\gamma _{i}\otimes v_{s^{\ast }})w_{st}^{\prime }e_{is^{\ast
}}e_{it}e_{is}\overset{p}{\rightarrow }0.
\end{equation*}%
As before, we look at the expectation of the square of the sum above, which
involves six-dimensional summation over time indexes and two-dimensional
cross-sectional summation (over $i,j$) and is normalized by $N^{-2}T^{-3}$.
Due to time-series independence of $e_{it}$, the six-dimensional summation
over time indexes has mostly zeros and can be reduced to three-dimensional
summation over time indexes as the set $\{s_{1},t_{1},s_{1}^{\ast
},s_{2},t_{2},s_{2}^{\ast }\}$ should have any distinct index to appear at
least twice. If we consider the cases when $i=j,$ then all terms are bounded
above by a constant and the number of non-zero terms is $NT^{3}$; given the
normalization, this sum converges to zero. When we sum over $i\neq j$, the
only part of $e_{it}$ that yields a non-trivial effect is $\pi _{i}^{\prime
}f_{t}$; hence this sum is bounded by
\begin{align*}
& \frac{1}{N^{2}}\sum_{i,j=1}^{T}\Vert \gamma _{i}\Vert \Vert \gamma
_{j}\Vert \Vert \pi _{i}\Vert ^{3}\Vert \pi _{j}\Vert ^{3}\max_{1\leq
s,t,s^{\ast }\leq T}\mathbb{E}\left[ \Vert v_{s^{\ast }}\Vert ^{2}\Vert
w_{st}\Vert ^{2}\Vert f_{s^{\ast }}\Vert ^{2}\Vert f_{s}\Vert ^{2}\Vert
f_{t}\Vert ^{2}\right] \\
& \leq C\left( \frac{1}{N}\sum_{i=1}^{N}\Vert \gamma _{i}\Vert \Vert \pi
_{i}\Vert ^{3}\right) ^{2}\leq \frac{1}{N^{2}}\max_{1\leq i\leq N}\Vert
\gamma _{i}\Vert ^{2}\max_{1\leq i\leq N}\Vert \pi _{i}\Vert
^{4}\sum_{i=1}^{N}\Vert \pi _{i}\Vert ^{2}\rightarrow 0.
\end{align*}%
This ends the proof of Theorem \ref{th: covariance for HC}. $\blacksquare $

\newpage

\section*{REFERENCES}

\begin{hangparas}{.2in}{1}

\noindent Anatolyev, S. \& A. Mikusheva (2018) Factor models with many
assets: strong factors, weak factors, and the two-pass procedure.\
Manuscript, CERGE-EI and MIT. Available as paper 1807.04094 on \texttt{%
arXiv.org}.\smallskip

\noindent Ando, T. \& J. Bai (2015) Asset pricing with a general multifactor
structure. \textit{Journal of Financial Econometrics} 13, 556--604.\smallskip

\noindent Bai, J. (2009) Panel data models with interactive fixed effects.
\textit{Econometrica} 77, 1229--1279.\smallskip

\noindent Bai, J. \& S. Ng (2006) Confidence intervals for diffusion index
forecast and inference with factor-augmented regressions. \textit{%
Econometrica} 74, 1133--1155.\smallskip

\noindent Bai, J. \& S. Ng (2010) Instrumental variable estimation in a data
rich environment. \textit{Econometric Theory} 26, 1577--1606.\smallskip

\noindent Bhansali, R.J.,  L. Giraitis \& P.S. Kokoszka (2007) Convergence of quadratic forms
with nonvanishing diagonal. \textit{Statistics \& Probability Letters} 77, 726--734.\smallskip

\noindent Cattaneo, M.D., R.K. Crump \& M. Jansson (2014a) Bootstrapping
density-weighted average derivatives. \textit{Econometric Theory} 30(6),
1135--1164.\smallskip

\noindent Cattaneo, M.D., R.K. Crump \& M. Jansson (2014b) Small bandwidth
asymptotics for density-weighted average derivatives. \textit{Econometric
Theory} 30(1), 176--200.\smallskip

\noindent Cattaneo, M.D., M. Jansson \& W.K. Newey (2018) Alternative
asymptotics and the partially linear model with many regressors. \textit{%
Econometric Theory} 34(2), 277--301.\smallskip

\noindent Chao, J.C., N.R. Swanson, J.A. Hausman, W.K. Newey \& T. Woutersen
(2012) Asymptotic distribution of JIVE in a heteroskedastic IV regression
with many instruments.\ \textit{Econometric Theory} 28, 42--86.\smallskip

\noindent Fama, E.F. \& J. MacBeth (1973) Risk, return and equilibrium:
Empirical tests.\ \textit{Journal of Political Economy} 81,
607--636.\smallskip

\noindent Giraitis, L., H.L. Koul \& D. Surgailis (2012)
\textit{Large Sample Inference for Long Memory Processes}.
Imperial College Press.\smallskip

\noindent Hagedorn, M., I. Manovskii \& K. Mitman (2015) The impact of
unemployment benefit extensions on employment: The 2014 employment miracle?
NBER working paper 20884.\smallskip

\noindent Hahn, J., G.M. Kuersteiner \& M. Mazzocco (2020) Central limit
theory for combined cross-section and time series. Manuscript, UCLA and
University of Maryland. Available as paper 1610.01697 on \texttt{arXiv.org}%
.\smallskip

\noindent Hausman, J.A., W.K. Newey, T. Woutersen, J.C. Chao \& N.R. Swanson
(2012) Instrumental variable estimation with heteroskedasticity and many
instruments.\ \textit{Quantitative Economics} 3, 211--255.\smallskip

\noindent Heyde, C. \& B. Brown (1970) On the departure from normality of a
certain class of martingales.\ \textit{Annals of Mathematical Statistics}
41, 2161--2165.\smallskip

\noindent de Jong, P. (1987) A central limit theorem for generalized
quadratic forms.\ \textit{Probability Theory and Related Fields} 75,
261--277.\smallskip

\noindent Kleibergen, F. \& Z. Zhan (2015) Unexplained factors and their
effects on second pass R-squared's.\ \textit{Journal of Econometrics} 189,
101--116.\smallskip

\noindent Kuersteiner, G.M. \& I.R. Prucha (2013) Limit theory for panel
data models with cross sectional dependence and sequential exogeneity.\
\textit{Journal of Econometrics} 174, 107--126.\smallskip

\noindent Kuersteiner, G.M. \& I.R. Prucha (2020) \textquotedblleft Dynamic
spatial panel models: Networks, common shocks, and sequential exogeneity.\
Forthcoming in \textit{Econometrica}. Available as paper 1802.01755 on \texttt{%
arXiv.org}.\smallskip

\noindent Onatski, A. (2012): Asymptotics of the principal components
estimator of large factor models with weakly influential factors. \textit{%
Journal of Econometrics} 168, 244--258. \smallskip

\noindent Pesaran, M.H. (2006) Estimation and inference in large
heterogeneous panels with a multifactor error structure. \textit{Econometrica%
} 74, 967--1012.\smallskip

\noindent Pesaran, M.H. \& T. Yamagata (2018) Testing for alpha in linear
factor pricing models with a large number of securities. Manuscript,
University of Southern California.\smallskip

\noindent Rotar', V.I. (1973) Some limit theorems for polynomials of second
degree. \textit{Theory of Probability and Its Applications} 18,
499--507.\smallskip

\noindent Sarto A.P. (2018) Recovering macro elasticities from regional
data. Manuscript, MIT.\smallskip

\noindent Serrato J.C.S. \& P. Wingender (2016) Estimating local fiscal
multipliers.\ NBER working paper 22425.\smallskip

\noindent Shanken, J. (1992) On the estimation of beta-pricing models.\
\textit{Review of Financial Studies} 5, 1--33.\smallskip

\noindent S\o lvsten, M. (2020) Robust estimation with many instruments.\
\textit{Journal of Econometrics} 214, 495--512.

\end{hangparas}

\end{document}